\documentclass[amsmath,amssymb,showkeys, showpacs,superscriptaddress]{revtex4}
\usepackage{graphicx}
\usepackage{latexsym}
\begin{document}

\hspace*{5 in}CUQM-148\\
\vspace*{0.4 in}
\title{Polynomial solutions for a class of second-order linear differential equations}

\author{Nasser Saad}
\email{nsaad@upei.ca}
\affiliation{Department of Mathematics and Statistics,
University of Prince Edward Island, 550 University Avenue,
Charlottetown, PEI, Canada C1A 4P3.}
\author{Richard L. Hall}
\email{rhall@mathstat.concordia.ca}
\affiliation{Department of Mathematics and Statistics, Concordia University,
1455 de Maisonneuve Boulevard West, Montr\'eal,
Qu\'ebec, Canada H3G 1M8.}
\author{Victoria A. Trenton}
\email{vtrenton@upei.ca}
\affiliation{Department of Mathematics and Statistics,
University of Prince Edward Island, 550 University Avenue,
Charlottetown, PEI, Canada C1A 4P3.}
\begin{abstract}
\noindent \textbf{Abstract:} We analyze the polynomial solutions of the linear differential equation 
$p_2(x)y''+p_1(x)y'+p_0(x)y=0$ where $p_j(x)$ is a $j^{\rm th}$-degree polynomial. We discuss all the possible polynomial solutions and their dependence on the parameters of the polynomials $p_j(x)$. Special cases are related to known differential equations of mathematical physics. Classes of new soluble problems are exhibited. General results are obtained for weight functions and orthogonality relations.

\end{abstract}

\maketitle
\section{Introduction}\label{intro}

\noindent The study of many physical phenomena can be reduced to the analysis of certain differential equations: these, in turn, demand insight for their solutions  \cite{bhat,bose,chihara,aim2,Dab,dekar,de,infeld,ismail,levai,nu,midya}. One of the simplest and most widely used linear differential equation is given \cite{ismail,nu,brenke,Bochner,hild,koepf1,koepf2,littlejohn,Routh,aim1} by
\begin{equation}\label{eq1}
(a_{2,0}x^2+a_{2,1}x+a_{2,2})y''+(a_{1,0}x+a_{1,1})y'-\tau_{0,0}y=0,\quad\quad '\equiv {d/dx},
\end{equation}
where the parameters $a_{j,k}, j=0,1,2, k=0,1,2$ are real constants and $\tau_{0,0}$ is a function of some non-negative integer $n$. It is known that the differential equation (\ref{eq1}) has an $n$-degree polynomial solution if
\begin{equation}\label{eq2}
\tau_{0,0}=n(n-1)a_{2,0}+na_{1,0},\quad n=0,1,2,\dots.
\end{equation}
The proof of this fact follows  by substituting $y(x)=\sum_{k=0}^nc_kx^k$ in  equation (\ref{eq1}) and differentiating the resulting equation $n$ times to obtain the condition (\ref{eq2}). On the other hand, if a differential equation has the form
\begin{equation}\label{eq3}
(a_{2,0}x^2+a_{2,1}x+a_{2,2})y''+(a_{1,0}x+a_{1,1})y'-\left[n(n-1)a_{2,0}+na_{1,0}\right]y=0,
\end{equation}
where $a_{2,0}^2+a_{1,0}^2\neq0$, then one solution of this differential equation is a polynomial of degree at most $n$. The proof of this claim is as follows. We first write equation (\ref{eq3}) as
\begin{equation}\label{eq4}
y''=\lambda_0(x) y'+s_0(x)y, 
\end{equation}
where
\begin{equation}\label{eq5}
\lambda_0(x)=-{(a_{1,0}x+a_{1,1})\over(a_{2,0}x^2+a_{2,1}x+a_{2,2})}\quad\mbox{and}\quad 
s_0(x)={n(n-1)a_{2,0}+na_{1,0}\over (a_{2,0}x^2+a_{2,1}x+a_{2,2})}.
\end{equation}
We then differentiate equation (\ref{eq4}) $n-1$ and $n$ times respectively to obtain
\begin{equation}\label{eq6}
y^{(n+1)}=\lambda_{n-1}(x) y'+s_{n-1}(x)y\quad\mbox{and}\quad
y^{(n+2)}=\lambda_{n}(x) y'+s_{n}(x)y,
\end{equation}
where
\begin{equation}\label{eq7}
\lambda_{n}(x)=\lambda'_{n-1}(x)+s_{n-1}(x)+\lambda_0(x)\lambda_{n-1}(x)\quad\mbox{and}\quad 
s_{n}(x)=s_{n-1}'(x)+s_0(x)\lambda_{n-1}(x),\quad n=1,2,3,\dots
\end{equation}
Now multiply the first equation of (\ref{eq6}) by $\lambda_{n-1}(x)$ and the second equation by $\lambda_n(x)$, then subtract the resulting equations to obtain
\begin{equation}\label{eq8}
\lambda_{n-1}y^{(n+2)}-\lambda_n y^{(n+1)}=\delta_n y,\quad \delta_n=\lambda_n(x) s_{n-1}(x)-\lambda_{n-1}(x) s_n(x).
\end{equation}
For $\lambda_0(x)$ and $s_0(x)$ given by (\ref{eq5}), by induction we show that $\delta_n=0$ for all $n=1,2,\dots$: For $n=1$, a straightforward computation, using (\ref{eq5}), allows us to conclude that $\delta_1=0$; assuming $\delta_n=0$ we have for
\begin{eqnarray*}
\delta_{n+1}&=&\lambda_{n+1}s_n-\lambda_n s_{n+1}=(\lambda^\prime_n+s_n+\lambda_0\lambda_n)s_n-\lambda_n(s^\prime_{n}+s_0\lambda_{n})\\
&=&s_n^2\left[\left({\lambda_n \over s_n}\right)^\prime+1+\lambda_0\left({\lambda_n\over s_n}\right)-s_0\left({\lambda_n\over s_n}\right)^2\right]\\
&=&s_n^2\left[\left({\lambda_{n-1} \over s_{n-1}}\right)^\prime+1+\lambda_0\left({\lambda_{n-1}\over s_{n-1}}\right)-s_0\left({\lambda_{n-1}\over s_{n-1}}\right)^2\right],\quad \mbox{since~~ $\delta_n=0$}\\
&=&s_n^2\left[{\lambda^\prime_{n-1}\over s_{n-1}}
-{\lambda_{n-1}s^\prime_{n-1} \over s^2_{n-1}}
+1+\lambda_0\left({\lambda_{n-1}\over s_{n-1}}\right)-s_0\left({\lambda_{n-1}\over s_{n-1}}\right)^2\right]\\
&=&s_n^2\left[{\lambda^\prime_{n-1}+s_{n-1}+\lambda_0\lambda_{n-1}\over s_{n-1}}
-{\lambda_{n-1}(s^\prime_{n-1}+s_0\lambda_{n-1})\over s^2_{n-1}}\right]=s_n^2\left[{\lambda_n\over s_{n-1}}-{s_n\lambda_{n-1}\over s^2_{n-1}}\right]={s_n^2 \over s^2_{n-1}}\cdot \delta_n=0.
\end{eqnarray*}
Equation (\ref{eq8}) then implies
\begin{equation}\label{eq9}
y^{(n+1)}=C_1\exp\left({\int^x{\lambda_{n-1}(\tau)\over \lambda_n(\tau)}d\tau}\right),
\end{equation}
where $C_1$ is a constant. Thus, equation (\ref{eq6}) reduces to a first-order linear equation with a solution of the form
\begin{equation}\label{eq10}
y(x)=\exp\left(-{\int^x{s_{n-1}(\tau)\over \lambda_{n-1}(\tau)}d\tau}\right).
\end{equation}
If we differentiate (\ref{eq10}) and substitute $y' =-s_{n-1}(x)\cdot y/ \lambda_{n-1}(x) $ in equation (\ref{eq6}) once again, we obtain $y^{(n+1)}(x)=0$, i.e. $y(x)$ is a polynomial of at most $n$.
\vskip0.1true in

\noindent The differential equation (\ref{eq3})  represents the source of many well-established results in the realm of special functions and orthogonal polynomials \cite{brenke,Bochner,chihara,hild,ismail,koepf1,koepf2,littlejohn,nu,Routh,krall,andrews}. The classification of the standard orthogonal polynomials, for example, follows from the analysis of the polynomial solutions of this differential \cite{Bochner,ismail}. These results have found very many applications in diverse areas of mathematical and theoretical physics. For illustrative purposes we cite only a few references here: \cite{bagchi,bhat,bose,chhajilany,fryantsev,aim2,Dab,dekar,dw,de,infeld,kruglov,krylov,krylov2,levai,lyskova,midya,pasquine,tan,turbiner}.  Most of the results concerning the polynomial solutions of equation (\ref{eq1}) are obtained either as special cases of the parameters $a_{j,k}$ or through various transformations to reduce the complexity of dealing with differential equation directly. For example, for $a_{2,0}\neq 0$, with 
$$p=\mp {\sqrt{a_{2,1}^2-4a_{2,0}a_{2,2}}\over a_{2,0}},\quad q={-a_{2,1}\pm\sqrt{a_{2,1}^2-4a_{2,0}a_{2,2}}\over 2a_{2,0}},$$
the change of variable $x=pt+q$ reduces the differential equation (\ref{eq3}) to the classical hypergeometric differential equation 
\begin{equation}\label{eq11}
t(1-t){d^2y\over dt^2}-{(a_{1,0}(pt+q)+a_{1,1})\over pa_{2,0}}{dy\over dt}+{n(n-1)a_{2,0}+na_{1,0}\over a_{2,0}}y=0,
\end{equation}
and with
\begin{align}\label{eq12}
p=-{\sqrt{a_{2,1}^2-4a_{2,0}a_{2,2}}\over 2a_{2,0}},\quad q=-{a_{2,1}\over 2a_{2,0}},
\end{align}
the differential equation (\ref{eq3}) reduces to 
\begin{equation}\label{eq13}
(1-t^2){d^2y\over dt^2}-{(a_{1,0}(pt+q)+a_{1,1})\over pa_{2,0}}{dy\over dt}+{n(n-1)a_{2,0}+na_{1,0}\over a_{2,0}}y=0.
\end{equation}
However, such transformations do not always allow us to study all the possible cases. For example, the case of $a_{2,1}^2-4a_{2,0}a_{2,2}\leq 0$ or $a_{2,0}=0$. 
The present work focuses on studying the polynomial solutions of equation (\ref{eq3}) directly,  without the application of  any transformation or reduction process. In other words, we study the polynomial solutions in terms of the parameters $a_{j,k}$ themselves. To our knowledge, no such study is available in the vast literature on the subject.
\vskip0.1true in
\noindent In Section II, we present the most general results concerning the polynomial solutions of equation (\ref{eq3}) and their recurrence relation. Some notes on the proofs of these results are given in Appendix I. We also discuss all the possible polynomial solutions and their dependence on the parameters of the polynomials $a_{j,k}, j=0,1,2, k=0,1,2$. Special cases that are related to the classical differential equations of mathematical physics are also presented. In Section III, we report some new differential equations that depend on arbitrary function and we explicitly compute their exact solutions in the form of products of polynomials and exponential functions.
\section{Polynomial solutions of equation (\ref{eq1})}
\noindent{\bf Theorem 1.} \emph{
The second-order linear differential equation (\ref{eq1}) 
has a polynomial solution of degree $n\in \mathbb N$ (the set of nonnegative integers) if, for fixed $n$,
\begin{equation}\label{eq14}
\tau_{0,0}=n(n-1)~a_{2,0}+n~a_{1,0},\quad n=0,1,2,\dots,
\end{equation}
provided $a_{2,0}^2+a_{1,0}^2\neq 0$. The polynomial solutions $y_n\equiv y_n(x)$ satisfy the three-term recurrence relation:
\begin{align}\label{eq15}
&y_{n+2}(x)=\bigg[A_nx+B_n\bigg]y_{n+1}(x)+C_ny_n(x),\quad n\geq 0,
\end{align} 
where the coefficients $A_n,B_n$ and $C_n$ are given by
\begin{align*}
A_n&={((2n+1)a_{2,0}+a_{1,0})(2(n+1)a_{2,0}+a_{1,0})\over (na_{2,0}+a_{1,0})},\\
B_n&={((2n+1)a_{2,0}+a_{1,0})(2n(n+1)a_{2,0}a_{2,1}+2(n+1)a_{1,0}a_{2,1}-2a_{1,1}a_{2,0}+a_{1,0}a_{1,1})\over (na_{2,0}+a_{1,0})(2na_{2,0}+a_{1,0})},\\
C_n&={(n+1)(2(n+1)a_{2,0}+a_{1,0})((4a_{2,2}a_{2,0}^2-a_{2,0}a_{2,1}^2)n^2+
(4a_{2,0}a_{1,0}a_{2,2}-a_{1,0}a_{2,1}^2)n
+a_{1,0}^2a_{2,2}-a_{1,1}a_{1,0}a_{2,1}+a_{2,0}a_{1,1}^2)\over (na_{2,0}+a_{1,0})(2na_{2,0}+a_{1,0})},
\end{align*}
initiated with 
\begin{equation*}
y_0(x)=1,\quad y_1(x)=a_{1,0}x+a_{1,1}.
\end{equation*}
Further, the polynomial solutions are given explicitly for the case of $a_{2,1}^2-4a_{2,0}a_{2,2}\neq 0$ as
\begin{align}\label{eq16}
&y_n\left(\begin{array}{lll}
a_{2,0} & a_{2,1} & a_{2,2} \\
a_{1,0} & a_{1,1}& ~ \\
\end{array}\bigg|x\right)={(-1)^n(\sqrt{a_{2,1}^2 - 4a_{2,0}a_{2,2}})^n
\left({2a_{2,0}a_{1,1}-a_{1,0}a_{2,1}-a_{1,0}\sqrt{a_{2,1}^2-4a_{2,0}a_{2,2}}\over -2a_{2,0}\sqrt{a_{2,1}^2-4a_{2,0}a_{2,2}}}\right)_n}\notag\\
&\times
{}_2F_1
\left(\begin{array}{ll}
-n & n-1+{a_{1,0}\over a_{2,0}} \\
{2a_{2,0}a_{1,1}-a_{1,0}a_{2,1}-a_{1,0}\sqrt{a_{2,1}^2-4a_{2,0}a_{2,2}}\over -2a_{2,0}\sqrt{a_{2,1}^2-4a_{2,0}a_{2,2}}}& ~ \\
\end{array}\bigg|{a_{2,0}x\over \sqrt{a_{2,1}^2-4a_{2,0}a_{2,2}}}+{a_{2,1}+\sqrt{a_{2,1}^2-4a_{2,0}a_{2,2}}\over 2\sqrt{a_{2,1}^2-4a_{2,0}a_{2,2}}}\right),
\end{align}
and for the case of $a_{2,1}^2-4a_{2,0}a_{2,2}= 0$ the polynomial solutions are given by
\begin{align}\label{eq17}
&y_n\left(\begin{array}{lll}
a_{2,0} & a_{2,1} & {a_{2,1}^2\over 4a_{2,0}} \\
a_{1,0} & a_{1,1}& ~ \\
\end{array}\bigg|x\right)=\left({2a_{2,0}a_{1,1}-a_{1,0}a_{2,1}\over 2a_{2,0}}\right)^n
{}_2F_0\left(-n, n-1+{a_{1,0}\over a_{2,0}};-;-{2a_{2,0}(a_{2,0}x+{1\over 2}a_{2,1})\over 2a_{2,0}a_{1,1}-a_{1,0}a_{2,1}}\right),
\end{align}
in which case, the differential equation reduces to
\begin{equation}\label{eq18}
a_{2,0}\left(x+{a_{2,1}\over 2a_{2,0}}\right)^2y''+(a_{1,0}x+a_{1,1})y'-\tau_{0,0}y=0,   
\end{equation}
and the coefficients $A_n,B_n$ and $C_n$ of three-term recurrence relation, equation (\ref{eq16}), are then
\begin{align*}
A_n&={((2n+1)a_{2,0}+a_{1,0})(2(n+1)a_{2,0}+a_{1,0})\over (na_{2,0}+a_{1,0})},\\
B_n&={((2n+1)a_{2,0}+a_{1,0})(2n(n+1)a_{2,0}a_{2,1}+2(n+1)a_{1,0}a_{2,1}-2a_{1,1}a_{2,0}+a_{1,0}a_{1,1})\over (na_{2,0}+a_{1,0})(2na_{2,0}+a_{1,0})},\\
C_n&={(n+1)(2(n+1)a_{2,0}+a_{1,0})(a_{1,0}a_{2,1}-2a_{1,1}a_{2,0})^2\over 4a_{2,0}(na_{2,0}+a_{1,0})(2na_{2,0}+a_{1,0})}.
\end{align*}
Here the functions ${}_2F_1$ and ${}_2F_0$ are special cases of the generalized hypergeometric function ${}_pF_q$ with $p$ numerator parameters $\alpha_j~(j = 1,\dots, p)$ and $q$ denominator 
$\beta_j\neq 0,-1,-2,\dots, j=1,\dots,q$ that is defined by \cite{andrews} as
\begin{equation}\label{eq19}
{}_pF_q\left(\begin{array}{lll}
\alpha_1 &\dots&\alpha_p\\
\beta_1 & \dots&\beta_q \\
\end{array}\bigg|x\right)=\sum_{k=0}^\infty 
{(\alpha_1)_k\dots (\alpha_p)_k\over (\beta_1)_k\dots(\beta_q)_k}{x^k\over k!},
\end{equation}
and $(\lambda)_k$ denotes the Pochhammer symbol defined, in terms of Gamma functions, by
$$(\lambda)_k= {\Gamma(\lambda+k)\over\Gamma(\lambda)}=
  \lambda(\lambda+1)(\lambda+2)\dots(\lambda+k-1). 
$$ 
For certain values of the constant parameters $a_{k,j}$, these polynomials are orthogonal  under a weight function given by the Pearson equation \cite{ismail}
\begin{align}\label{eq20}
W\left(\begin{array}{lll}
a_{2,2} & a_{2,1} & a_{2,2} \\
a_{1,0} & a_{1,1}& ~ \\
\end{array}\bigg|x\right)&={1\over a_{2,0} x^2+a_{2,1} x+a_{2,2}}
\exp\left({\int^x {a_{1,0} \tau+a_{1,1}\over a_{2,0} \tau^2+a_{2,1}\tau+a_{2,2}}}d\tau\right), 
 \end{align}
where $W$ is non-negative in the interval of definition (the interval could reach to infinity in either or both directions).}
\vskip0.1true in
\noindent A note on the proof of this theorem can be found in Appendix I.
\section*{Case I: $a_{2,0}=0$}
\vskip0.1true in
\noindent In this case the differential equation (\ref{eq3}) reads
\begin{equation}\label{eq21}
(a_{2,1}x+a_{2,2})~y_n^{\prime \prime}+(a_{1,0}x+a_{1,1})~y_n'-n~a_{1,0}~y_n=0,\quad n=0,1,2,\dots
\end{equation}
and the recurrence relation reduces, in the case of $a_{2,1}a_{1,1} - a_{2,2}a_{1,0}\neq 0$, to
\begin{equation}\label{eq22}
\left\{ \begin{array}{l}
 y_{n+2}=\left(a_{1,0}x+2(n+1)a_{2,1}+a_{1,1}\right)~y_{n+1}-
(n+1)(a_{2,1}^2n+a_{1,1}a_{2,1}-a_{1,0}a_{2,2})~y_n, \\  
  y_0=1,\quad y_1=a_{1,0}x+a_{1,1}\quad\quad (n=0,1,2,\dots),
       \end{array} \right.
\end{equation}
Meanwhile, for $a_{2,1}a_{1,1} - a_{2,2}a_{1,0}= 0$, the recurrence relation reads 
\begin{equation}\label{eq23}
\left\{ \begin{array}{l}
 y_{n+2}=\left(a_{1,0}x+2(n+1)a_{2,1}+{a_{2,2}a_{1,0}\over a_{2,1}}\right)~y_{n+1}-
n(n+1)~a_{2,1}^2~y_n, \\ 
  y_0=1,\quad y_1=a_{1,0}x+{a_{2,2}a_{1,0}\over a_{2,1}}\quad\quad (n=0,1,2,\dots).
       \end{array} \right.
\end{equation}
The exact polynomial solutions can be expressed in terms of hypergeometric functions as follows: 
\vskip0.1true in
\noindent $\bullet$ For $a_{2,1}a_{1,1} - a_{2,2}a_{1,0}\neq 0$
\begin{align}\label{eq24}
y_n\left(\begin{array}{lll}
0 & a_{2,1} & a_{2,2} \\
a_{1,0} & a_{1,1}& ~ \\
\end{array}\bigg|x\right)
&=a_{2,1}^n\left({a_{2,1}a_{1,1} - a_{2,2}a_{1,0}\over a_{2,1}^2}\right)_n
{}_1F_1\left(-n;{a_{2,1}a_{1,1} - a_{2,2}a_{1,0}\over a_{2,1}^2}
;-{a_{1,0}\over a_{2,1}}x - {a_{2,2}a_{1,0}\over a_{2,1}^2}\right).
\end{align}
This can be obtained as a limiting case of equation (\ref{eq16}) for $a_{2,0}\rightarrow 0$ as follows. From the series representation of the Gauss hypergeometric function ${}_2F_1$, see equation (\ref{eq19}), we may write equation (\ref{eq16}) as
\begin{align*}
&y_n\left(\begin{array}{lll}
a_{2,0} & a_{2,1} & a_{2,2} \\
a_{1,0} & a_{1,1}& ~ \\
\end{array}\bigg|x\right)={(-1)^n
\left(\sqrt{a_{2,1}^2 - 4a_{2,0}a_{2,2}}\right)^n
\left(-{a_{1,1}\over \sqrt{a_{2,1}^2-4a_{2,0}a_{2,2}}}
+{a_{1,0}\over a_{2,0}} \left(
{a_{2,1}+
\sqrt{a_{2,1}^2-4a_{2,0}a_{2,2}}\over 2\sqrt{a_{2,1}^2-4a_{2,0}a_{2,2}}}\right)\right)_n}\notag\\
&\times
\sum_{k=0}^n{(-n)_k(n-1+{a_{1,0}\over a_{2,0}})_k\over k!\left(-{a_{1,1}\over \sqrt{a_{2,1}^2-4a_{2,0}a_{2,2}}}
+{a_{1,0}\over a_{2,0}} \left(
{a_{2,1}+
\sqrt{a_{2,1}^2-4a_{2,0}a_{2,2}}\over 2\sqrt{a_{2,1}^2-4a_{2,0}a_{2,2}}}\right)\right)_k }
\left({a_{2,0}x\over \sqrt{a_{2,1}^2-4a_{2,0}a_{2,2}}}+{a_{2,1}+\sqrt{a_{2,1}^2-4a_{2,0}a_{2,2}}\over 2\sqrt{a_{2,1}^2-4a_{2,0}a_{2,2}}}\right)^k.
\end{align*}
By rationalizing the factor $a_{2,1}+
\sqrt{a_{2,1}^2-4a_{2,0}a_{2,2}}$, we obtain
\begin{align*}
y_n\left(\begin{array}{lll}
a_{2,0} & a_{2,1} & a_{2,2} \\
a_{1,0} & a_{1,1}& ~ \\
\end{array}\bigg|x\right)&=(-1)^n
\left(\sqrt{a_{2,1}^2 - 4a_{2,0}a_{2,2}}\right)^n
\left({-a_{1,1}\over \sqrt{a_{2,1}^2-4a_{2,0}a_{2,2}}}
+
{2a_{1,0}a_{2,2}\over \sqrt{a_{2,1}^2-4a_{2,0}a_{2,2}}(a_{2,1}-\sqrt{a_{2,1}^2-4a_{2,0}a_{2,2}})}\right)_n\notag\\
&\times
\sum_{k=0}^n{(-n)_k(n-1+{a_{1,0}\over a_{2,0}})_k~a_{2,0}^k
\left({x\over \sqrt{a_{2,1}^2-4a_{2,0}a_{2,2}}}+{2a_{2,2}\over \sqrt{a_{2,1}^2-4a_{2,0}a_{2,2}}(a_{2,1}-\sqrt{a_{2,1}^2-4a_{2,0}a_{2,2}}}\right)^k\over k!
\left(-{a_{1,1}\over \sqrt{a_{2,1}^2-4a_{2,0}a_{2,2}}}
+
{2a_{1,0}a_{2,2}\over \sqrt{a_{2,1}^2-4a_{2,0}a_{2,2}}(a_{2,1}-\sqrt{a_{2,1}^2-4a_{2,0}a_{2,2}})}\right)_k }.
\end{align*}
We perform the limit operation, as $a_{2,0}$ approaches zero, using the identity
$\lim_{a_{2,0}\rightarrow 0} a_{2.,0}^k\left(n-1+{a_{1,0}\over a_{2,0}}\right)_k=a_{1,0}^k,
$ as well as the fact that for $a_{2,1}<0$ we may write $\sqrt{a_{2,1}^2}=|a_{2,1}|=-a_{2,1}$, to obtain 
\begin{align*}
&y_n\left(\begin{array}{lll}
0 & a_{2,1} & a_{2,2} \\
a_{1,0} & a_{1,1}& ~ \\
\end{array}\bigg|x\right)=(-1)^n
(-a_{2,1})^n
\left({a_{1,1}\over a_{2,1}}
-
{a_{1,0}a_{2,2}\over a_{2,1}^2}\right)_n
\sum_{k=0}^n{(-n)_ka_{1,0}^k\over k!
\left({a_{1,1}\over a_{2,1}}
-
{a_{1,0}a_{2,2}\over a_{2,1}^2}\right)_k}
\left(-{x\over a_{2,1}}-{a_{2,2}\over a_{2,1}^2}\right)^k,
\end{align*}
which reduces to equation (\ref{eq24}) by use of the series representation of the confluent hypergeometric function ${}_1F_1$.
\vskip0.1true in
\noindent $\bullet$ For $a_{2,1}a_{1,1} - a_{2,2}a_{1,0}= 0$, the polynomial solutions read
\begin{align}\label{eq25}
y_n\left(\begin{array}{lll}
0 & a_{2,1} & a_{2,2} \\
a_{1,0} & {a_{2,2}a_{1,0}\over a_{2,1}}& ~ \\
\end{array}\bigg|x\right)
&=\left\{ \begin{array}{ll}
 1 &\mbox{if $n=0$,} \\ 
 a_{2,1}^n~n!\left({a_{1,0}\over a_{2,1}}x + {a_{2,2}a_{1,0}\over a_{2,1}^2}\right)
{}_1F_1\left(1-n;2;-{a_{1,0}\over a_{2,1}}x - {a_{2,2}a_{1,0}\over a_{2,1}^2}\right)&\mbox{if $n\geq 1$}.
       \end{array} \right.
\end{align}
To obtain this equation from (\ref{eq24}) we use the series representation of the confluent hypergeometric functions (see equation (\ref{eq19})) and we write
\begin{align*}
y_n\left(\begin{array}{lll}
0 & a_{2,1} & a_{2,2} \\
a_{1,0} & a_{1,1}& ~ \\
\end{array}\bigg|x\right)
&=a_{2,1}^n\sum_{k=0}^n{(-n)_k\left({a_{2,1}a_{1,1} - a_{2,2}a_{1,0}\over a_{2,1}^2}\right)_n\over ({a_{2,1}a_{1,1} - a_{2,2}a_{1,0}\over a_{2,1}^2})_k~k!}\left(-{a_{1,0}\over a_{2,1}}x - {a_{2,2}a_{1,0}\over a_{2,1}^2}\right)^k.
\end{align*}
By means of the identity
${(\alpha)_n/(\alpha)_k}=(\alpha+k)_{n-k}$
we have
\begin{align*}
y_n\left(\begin{array}{lll}
0 & a_{2,1} & a_{2,2} \\
a_{1,0} & a_{1,1}& ~ \\
\end{array}\bigg|x\right)
&=a_{2,1}^n\sum_{k=0}^n{(-n)_k\left({a_{2,1}a_{1,1} - a_{2,2}a_{1,0}\over a_{2,1}^2}+k\right)_{n-k}\over k!}\left(-{a_{1,0}\over a_{2,1}}x - {a_{2,2}a_{1,0}\over a_{2,1}^2}\right)^k.
\end{align*}
Thus for $a_{2,1}a_{1,1} - a_{2,2}a_{1,0}\rightarrow 0$, we obtain
\begin{align*}
y_n\left(\begin{array}{lll}
0 & a_{2,1} & a_{2,2} \\
a_{1,0} & a_{1,1}& ~ \\
\end{array}\bigg|x\right)
&=a_{2,1}^n\sum_{k=0}^n{(-n)_k\left(k\right)_{n-k}\over k!}\left(-{a_{1,0}\over a_{2,1}}x - {a_{2,2}a_{1,0}\over a_{2,1}^2}\right)^k,
\end{align*}
if $n=0$, using the identity $(0)_0=1$, we have
$$
y_0\left(\begin{array}{lll}
0 & a_{2,1} & a_{2,2} \\
a_{1,0} & a_{1,1}& ~ \\
\end{array}\bigg|x\right)
=1,
$$ 
while for $n\geq 1$ where $0\leq k\leq n$ and using $(k)_{n-k}=\Gamma(n)/\Gamma(k),$ we have
\begin{align*}
y_n\left(\begin{array}{lll}
0 & a_{2,1} & a_{2,2} \\
a_{1,0} & a_{1,1}& ~ \\
\end{array}\bigg|x\right)
&=-n!~a_{2,1}^n\left(-{a_{1,0}\over a_{2,1}}x - {a_{2,2}a_{1,0}\over a_{2,1}^2}\right)\sum_{k=0}^{n-1}{(-n+1)_k\over (2)_k~k!}\left(-{a_{1,0}\over a_{2,1}}x - {a_{2,2}a_{1,0}\over a_{2,1}^2}\right)^{k}\\
&=n!~a_{2,1}^n\left({a_{1,0}\over a_{2,1}}x + {a_{2,2}a_{1,0}\over a_{2,1}^2}\right){}_1F_1\left(-n+1;2;-{a_{1,0}\over a_{2,1}}x - {a_{2,2}a_{1,0}\over a_{2,1}^2}\right).\quad 
\end{align*}
The numerical coefficients of the polynomial solutions can be written for $a_{2,1}a_{1,1} - a_{2,2}a_{1,0}\neq 0$ as
\begin{equation}\label{eq26}
\left\{ \begin{array}{l}
 y_0=1 \\  
 y_1=a_{1,0}x+a_{1,1}\\ 
 y_2=a_{1,0}^2x^2+2a_{1,0}(a_{2,1}+a_{1,1})x+a_{1,1}^2+a_{1,0}a_{2,2}+a_{1,1}a_{2,1}\\ 
 y_3=a_{1,0}^3x^3+3a_{1,0}^2(2a_{2,1}+a_{1,1})x^2+3a_{1,0}((2a_{2,1}+a_{1,1})(a_{1,1}+a_{2,1})+a_{1,0}a_{2,2})x\\
 \quad~~+3a_{1,1}a_{1,0}a_{2,2}+a_{1,1}^3+3a_{2,1}a_{1,1}^2+4a_{2,1}a_{1,0}a_{2,2}+2a_{1,1}a_{2,1}^2\\ 
 y_4=a_{1,0}^4x^4+4a_{1,0}^3(3a_{2,1}+a_{1,1})x^3
 +6a_{1,0}^2((3a_{2,1}+a_{1,1})(a_{1,1}+2a_{2,1})+a_{1,0}a_{2,2})x^2\\
\quad~~+4a_{1,0}(a_{1,1}^3+6a_{2,1}a_{1,1}^2+11a_{1,1}a_{2,1}^2+3a_{1,0}a_{1,1}a_{2,2}+6a_{2,1}^3+7a_{1,0}a_{2,1}a_{2,2})x \\
\quad~~+6a_{1,0}a_{2,2}a_{1,1}^2+22a_{1,0}a_{2,2}a_{1,1}a_{2,1}+11a_{1,1}^2a_{2,1}^2+6a_{1,1}a_{2,1}^3+6a_{1,1}^3a_{2,1}+3a_{1,0}^2a_{2,2}^2+18a_{1,0}a_{2,2}a_{2,1}^2+a_{1,1}^4\\
\dots
             \end{array} \right.
\end{equation}
while for $a_{2,1}a_{1,1} - a_{2,2}a_{1,0}=0$, the numerical coefficients are obtained by replacing $a_{1,1}$ with $a_{2,2}a_{1,0}/a_{2,1}$. For the weight function and orthogonal polynomials, we have for $a_{2,0}=0$ and $a_{2,1}\neq 0$, the following two cases: 
\vskip0.1true in
\noindent $\bullet$ The constant $a_{2,1}>0$, the Pearson equation (\ref{eq20}) for the weight function yields 
  \begin{align}\label{eq27}
W_1\left(\begin{array}{lll}
0 & a_{2,1} & a_{2,2} \\
a_{1,0} & a_{1,1}& ~ \\
\end{array}\bigg|x\right)
&=
(a_{2,2} + a_{2,1} x)^{ {a_{1,1} a_{2,1} - a_{1,0} a_{2,2}\over a_{2,1}^2}-1}e^{{a_{1,0}\over a_{2,1}} x},\quad x\in (-{a_{2,2}/a_{2,1}},\infty)\end{align}
which required $a_{1,1} > a_{2,2}a_{1,0}/a_{2,1}$ and $ 
a_{1,0}<0$. In this case, the orthogonality condition reads
\begin{align}\label{eq28}
\int\limits_{-{a_{2,2}\over a_{2,1}}}^\infty 
y_n(x)y_m(x)
W_1(x)~dx= \left\{ \begin{array}{ll}
 0, &\mbox{ if $m\neq n$}, \\
n!~a_{2,1}^{{2n}-1}~
\Gamma({a_{2,1}a_{1,1} - a_{2,2}a_{1,0}\over a_{2,1}^2}+n)
~e^{-{a_{1,0}a_{2,2}\over a_{2,1}^2}}~\left(-{ a_{2,1}^2\over a_{1,0}}\right)^{{a_{2,1}a_{1,1} - a_{2,2}a_{1,0}\over a_{2,1}^2}},
 &\mbox{if $m=n$,}
       \end{array} \right.
\end{align}
where $y_n(x)$ is given by (\ref{eq24}).
To prove this orthogonality relation, we start with the series representation of the confluent hypergeometric function, see (\ref{eq19}), and write
\begin{align*}
&\int\limits_{-{a_{2,2}\over a_{2,1}}}^\infty 
y_n(x)y_m(x)
W_1(x)~dx=a_{2,1}^{n+m}
\left(\mu\right)_n\left(\mu\right)_m\sum_{k=0}^n\sum_{j=0}^m{(-n)_k(-m)_j\over (\mu)_k(\mu)_j~k!~j!}\left(-{a_{1,0}\over a_{2,1}^2}\right)^{k+j}I_{nm}
\end{align*}
where, for simplicity,
$\mu={(a_{2,1}a_{1,1} - a_{2,2}a_{1,0})/a_{2,1}^2}$ and
\begin{align*}
I_{nm}=\int\limits_{-{a_{2,2}\over a_{2,1}}}^\infty 
(a_{2,2} + a_{2,1} x)^{k+j+\mu-1}e^{{a_{1,0}\over a_{2,1}} x}~dx~~{\overset {(z=x+a_{2,2}/a_{2,1})}=}~ a_{2,1}^{k+j+\mu-1}e^{-{a_{1,0}a_{2,2}\over a_{2,1}^2}}\int\limits_0^\infty 
z^{k+j+\mu-1}
e^{{a_{1,0}\over a_{2,1}}z}~dz.
\end{align*}
Using the integral representation of Gamma function, we have for $k+j+\mu>0$ and $a_{1,0}/a_{2,1}<0$,
\begin{align*}
I_{nm}&={e^{-{a_{1,0}a_{2,2}\over a_{2,1}^2}}\over a_{2,1}\left(-{a_{1,0}\over a_{2,1}^2}\right)^{k+j+\mu}}
\left(k+\mu\right)_j\left(\mu\right)_k\Gamma\left(\mu\right)\\
\end{align*}
where we have used the identity $\Gamma(\mu+k+j)=(\mu+k)_j(\mu)_j\Gamma(\mu).$
Thus, we obtain
\begin{align*}
\int\limits_{-{a_{2,2}\over a_{2,1}}}^\infty 
y_n(x)y_m(x)
W_1(x)~dx&={\Gamma\left(\mu\right)
a_{2,1}^{n+m}
\left(\mu\right)_n\left(\mu\right)_me^{-{a_{1,0}a_{2,2}\over a_{2,1}^2}}\over a_{2,1}\left(-{a_{1,0}\over a_{2,1}^2}\right)^{\mu}}
\sum_{k=0}^n\sum_{j=0}^m{(-n)_k(-m)_j\left(k+\mu\right)_j\left(\mu\right)_k\over (\mu)_k(\mu)_j~k!~j!}
\\
&={\Gamma\left(\mu\right)
a_{2,1}^{n+m}
\left(\mu\right)_n\left(\mu\right)_m\left(-{a_{1,0}\over a_{2,1}^2}\right)^{-\mu}e^{-{a_{1,0}a_{2,2}\over a_{2,1}^2}}\over a_{2,1}}
\sum_{k=0}^n{(-n)_k\over k!}{}_2F_1(-m,k+\mu;\mu;1)\\
&=a_{2,1}^{n+m}\left(\mu\right)_n\left(-{a_{1,0}\over a_{2,1}^2}\right)^{-\mu}{e^{-{a_{1,0}a_{2,2}\over a_{2,1}^2}}\over a_{2,1}}\Gamma\left(\mu\right)
\sum_{k=0}^n{(-n)_k(-k)_m\over k!},
\end{align*}
in which we have used the Chu-Vandermonde summation formula
${}_2F_1(-n,\alpha;\beta;1)={(\beta-\alpha)_n/(\beta)_n}$.
Since
$\sum_{k=0}^n(-n)_k{(-k)_m}/k!=n!~\delta_{nm},
$
where $\delta_{nm}=0$ if $n\neq m$ or $\delta_{nm}=1$ if $n= m$, we finally have the orthogonality relation 
\begin{align*}
\int\limits_{-{a_{2,2}\over a_{2,1}}}^\infty 
y_n(x)y_m(x)
W_1(x)dx=a_{2,1}^{2n-1}{n!e^{-{a_{1,0}a_{2,2}\over a_{2,1}^2}}}
\Gamma\left(n+{a_{2,1}a_{1,1} - a_{2,2}a_{1,0}\over a_{2,1}^2}\right)
\left(-{a_{1,0}\over a_{2,1}^2}\right)^{-{a_{2,1}a_{1,1} - a_{2,2}a_{1,0}\over a_{2,1}^2}}\delta_{nm}.
\end{align*}
\vskip0.1true in
\noindent$\bullet$  The constant $a_{2,1}<0$, we have for  $a_{2,2}>0$,
$a_{1,0}<0$   and $a_{1,1}<{a_{1,0}a_{2,2}/a_{2,1}}$, the orthogonality condition
\begin{align}\label{eq29}
\int\limits_{-\infty}^{-{a_{2,2}\over a_{2,1}}} y_n(x)y_m(x)W_1(x)
dx= \left\{ \begin{array}{ll}
 0, &\mbox{ if $m\neq n$}, \\ 
-n!~a_{2,1}^{{2n}-1}~
\Gamma({a_{2,1}a_{1,1} - a_{2,2}a_{1,0}\over a_{2,1}^2}+n)
~e^{-{a_{1,0}a_{2,2}\over a_{2,1}^2}}~\left(-{ a_{2,1}^2\over a_{1,0}}\right)^{{a_{2,1}a_{1,1} - a_{2,2}a_{1,0}\over a_{2,1}^2}},
 &\mbox{if $m=n$.}
       \end{array} \right.
\end{align}
The proof of this identity follows similarly to the proof of  (\ref{eq28}), therefore we omit it.
\section*{Case II: $a_{2,2}=0$}
\noindent In this case the differential equation (\ref{eq3}) reads
\begin{equation}\label{eq30}
x(a_{2,0}x+a_{2,1})~y^{\prime \prime}+(a_{1,0}x+a_{1,1})~y'-n((n-1)a_{2,0}+a_{1,0})~y=0,
\end{equation}
and the recurrence relation to generate the polynomial solutions is given by: For $a_{1,0}a_{2,1} - a_{2,0}a_{1,1}\neq 0$
\begin{align}\label{eq31}
y_{n+2}&=\bigg({((2n+1)a_{2,0}+a_{1,0})(2(n+1)a_{2,0}+a_{1,0})\over na_{2,0}+a_{1,0}}x\notag\\
&+
{((2n+1)a_{2,0}+a_{1,0})(2a_{2,0}a_{2,1}n^2+2a_{2,1}(a_{2,0}+a_{1,0})n+2a_{1,0}a_{2,1}-2a_{1,1}a_{2,0}+a_{1,0}a_{1,1})\over (na_{2,0}+a_{1,0})(2na_{2,0}+a_{1,0})}\bigg)~y_{n+1}\notag\\
&
-{(n+1)(2(n+1)a_{2,0}+a_{1,0})(a_{2,0}a_{2,1}^2n^2+a_{1,0}a_{2,1}^2n+a_{2,1}a_{1,1}a_{1,0}-a_{2,0}a_{1,1}^2)\over (na_{2,0}+a_{1,0})(2na_{2,0}+a_{1,0})}~y_n,
\end{align}
while for $a_{1,0}a_{2,1} - a_{2,0}a_{1,1}= 0$, it reads
\begin{align}\label{eq32}
y_{n+2}&=\bigg({((2n+1)a_{2,0}+a_{1,0})(2(n+1)a_{2,0}+a_{1,0})\over na_{2,0}+a_{1,0}}x+
{a_{2,1}((2n+1)a_{2,0}+a_{1,0})(2a_{2,0}^2n^2+2a_{2,0}(a_{2,0}+a_{1,0})n+a_{1,0}^2)\over a_{2,0}(na_{2,0}+a_{1,0})(2na_{2,0}+a_{1,0})}\bigg)~y_{n+1}\notag\\
&-{a_{2,1}^2n(n+1)(2(n+1)a_{2,0}+a_{1,0})\over (2na_{2,0}+a_{1,0})}~y_n,
\end{align}
where $y_0(x)=1,y_1(x)=a_{1,0}x+a_{1,1}$. In terms of the hypergeometric functions, the polynomial solutions in the case of $a_{1,0}a_{2,1} - a_{2,0}a_{1,1}\neq 0$ and $a_{2,1}>0$ are
\begin{align}\label{eq33}
y_n\left(\begin{array}{lll}
a_{2,0} & a_{2,1} & 0 \\
a_{1,0} & a_{1,1}& ~ \\
\end{array}\bigg|x\right)&={(-1)^na_{2,1}^n
\left({a_{1,0}a_{2,1}-a_{2,0}a_{1,1}\over a_{2,0}a_{2,1}}\right)_n}
{}_2F_1(-n,n-1+{a_{1,0}\over a_{2,0}} ;{a_{1,0}a_{2,1}-a_{2,0}a_{1,1}\over a_{2,0}a_{2,1}};{a_{2,0}\over a_{2,1}}x+1)
\end{align}
which can be easily obtained from equation (\ref{eq16}) as a limit case of $a_{2,2}\rightarrow 0$, while for $a_{1,0}a_{2,1} - a_{2,0}a_{1,1}= 0$, the polynomial solution is simplified to
\begin{align}\label{eq34}
y_n\left(\begin{array}{lll}
 a_{2,0} & a_{2,1} &0 \\
a_{1,0} & {a_{2,1}a_{1,0}\over a_{2,0}}& ~ \\
\end{array}\bigg|x\right)
&=\left\{ \begin{array}{ll}
 1, &\mbox{if $n=0$,} \\
 (-1)^{n+1} n!~ a_{2,1}^n\left(n-1+{a_{1,0}\over a_{2,0}}\right)\left({a_{2,0}\over a_{2,1}}x+1\right)
{}_2F_1\left(1-n,n+{a_{1,0}\over a_{2,0}};2;{a_{2,0}\over a_{2,1}}x +1\right),&\mbox{if $n\geq 1$.}
       \end{array} \right.
\end{align}
To prove this identity, we use the series representation of the Gauss hypergeometric function (see equation (\ref{eq19})) to write
equation (\ref{eq33}), for $\mu= {(a_{1,0}a_{2,1} - a_{2,0}a_{1,1})/(a_{2,0}a_{2,1})}$, as
\begin{align*}
y_n\left(\begin{array}{lll}
a_{2,0} & a_{2,1} & 0 \\
a_{1,0} & a_{1,1}& ~ \\
\end{array}\bigg|x\right)
&=(-1)^na_{2,1}^n
\sum_{k=0}^n {(-n)_k\left(\mu+k\right)_{n-k}(n-1+{a_{1,0}\over a_{2,0}})_k\over 
k!}\left({a_{2,0}\over a_{2,1}}x +1\right)^k.
\end{align*}
Thus, as $a_{1,0}a_{2,1} - a_{2,0}a_{1,1}\rightarrow 0$, and since $(k)_{n-k}={\Gamma(n)/\Gamma(k)}$, we have for $n=0$
that \begin{align*}
y_n\left(\begin{array}{lll}
a_{2,0} & a_{2,1} & 0 \\
a_{1,0} & {a_{2,1}a_{1,0}\over a_{2,0}}& ~ \\
\end{array}\bigg|x\right)=1
\end{align*}
while for $n\geq 1$,
\begin{align*}
y_n\left(\begin{array}{lll}
a_{2,0} & a_{2,1} & 0 \\
a_{1,0} & {a_{2,1}a_{1,0}\over a_{2,0}}& ~ \\
\end{array}\bigg|x\right)
&=(-1)^{n+1}\Gamma(n+1)a_{2,1}^n(n-1+{a_{1,0}\over a_{2,0}})\left({a_{2,0}\over a_{2,1}}x +1\right)
\sum_{k=0}^{n-1} {(-n+1)_{k}(n+{a_{1,0}\over a_{2,0}})_{k}\over 
\Gamma(k+2)~k!}\left({a_{2,0}\over a_{2,1}}x +1\right)^{k}\\
&=(-1)^{n+1}~n!~a_{2,1}^n\left(n-1+{a_{1,0}\over a_{2,0}}\right)\left({a_{2,0}\over a_{2,1}}x +1\right)
{}_2F_1\left(-n+1, n+{a_{1,0}\over a_{2,0}};2;{a_{2,0}~x+ a_{2,1} \over a_{2,1}}\right).
\end{align*}
The numerical coefficients of these polynomial solutions for $a_{1,0}a_{2,1} - a_{2,0}a_{1,1}\neq 0$ are
\begin{equation}\label{eq35}
\left\{ \begin{array}{l}
 y_0=1 \\  
 y_1=a_{1,0}x+a_{1,1}\\ 
 y_2=(a_{1,0} + a_{2,0}) (a_{1,0} + 2 a_{2,0}) x^2+ 2 (a_{1,1}+a_{2,1}) (a_{1,0} + a_{2,0}) x+a_{1,1}(a_{1,1}+a_{2,1}) \\
 y_3=(a_{1,0} + 2 a_{2,0}) (a_{1,0} + 3 a_{2,0}) (a_{1,0} + 4 a_{2,0}) x^3+
 3 (a_{1,1}+2a_{2,1}) (a_{1,0} + 2 a_{2,0}) (a_{1,0} + 3 a_{2,0}) x^2\\
 +3(a_{2,1}+a_{1,1})( a_{1,1}+2a_{2,1})(a_{1,0} + 2 a_{2,0})x + a_{1,1}(a_{1,1}+a_{2,1})(a_{1,1}+2a_{2,1})\\ 
 y_4=(a_{1,0}+6a_{2,0})(a_{1,0}+5a_{2,0})(a_{1,0}+4a_{2,0})(a_{1,0}+3a_{2,0})x^4\\
+4(a_{1,1}+3a_{2,1})(a_{1,0}+5a_{2,0})(a_{1,0}+4a_{2,0})(a_{1,0}+3a_{2,0})x^3+6(a_{1,1}+3a_{2,1})(a_{1,1}+2a_{2,1})(a_{1,0}+3a_{2,0})(a_{1,0}+4a_{2,0})x^2\\
+4(a_{1,1}+3a_{2,1})(a_{1,1}+2a_{2,1})(a_{1,1}+a_{2,1})(a_{1,0}+3a_{2,0})x+ a_{1,1}(a_{1,1}+a_{2,1})(a_{1,1}+2a_{2,1})(a_{1,1}+3a_{2,1})\\ 
  \dots
             \end{array} \right.
\end{equation}
while for $a_{1,0}a_{2,1} - a_{2,0}a_{1,1}=0$, we obtain the numerical coefficient of the polynomial solutions by replacing $a_{1,1}$ with ${a_{1,0}a_{2,1}/ a_{2,0}}$. For the weight function and the orthogonally condition, we note that by using the Pearson equation (\ref{eq20}) we have 
\begin{align}\label{eq36}
W_2\left(\begin{array}{lll}
a_{2,0} & a_{2,1} & 0 \\
a_{1,0} & a_{1,1}& ~ \\
\end{array}\bigg|x\right)
&=
(a_{2,1} + a_{2,0} x)^{ {a_{1,0} \over a_{2,0}}-{a_{1,1}\over a_{2,1}}-1} x^{{a_{1,1}\over a_{2,1}}-1},
\end{align} 
where $a_{2,0}<0,~ a_{2,1}>0,a_{1,0}<0,~ 0<a_{1,1}<{a_{1,0}a_{2,1}/a_{2,0}}$,~ $x\in[0,-{a_{2,1}/a_{2,0}})$, we have the orthogonality relation
\begin{align}\label{eq37}
\int\limits_{0}^{{-{a_{2,1}\over a_{2,0}}}}
y_n(x)&y_m(x)W_2(x)dx= \left\{ \begin{array}{ll}
 0, &\mbox{\quad if $m\neq n$}, \\
n!~(-a_{2,0})^{-{a_{1,1}\over a_{2,1}}}~a_{2,1}^{2n-1+{a_{1,0}\over a_{2,0}}}{(a_{1,0}+(n-1)a_{2,0})\over (a_{1,0}+(2n-1)a_{2,0})}
{\Gamma(n+{a_{1,0}\over a_{2,0}}-{a_{1,1}\over a_{2,1}}) \Gamma(n+{a_{1,1}\over a_{2,1}})\over \Gamma(n+{a_{1,0}\over a_{2,0}})},&\mbox{\quad if $m=n$.}
       \end{array} \right.
\end{align}
where $y_n(x)$ is given by equation (\ref{eq33}). Denote $\mu={(a_{1,0}a_{2,1}-a_{2,0}a_{1,1})/(a_{2,0}a_{2,1})}$ we have using the series representation of the Gauss hypergeometric function, see equation (\ref{eq19}), for $y_n(x)$ as given by (\ref{eq33}), that
\begin{align*}
\int\limits_{0}^{{-{a_{2,1}\over a_{2,0}}}}
&y_n(x)y_m(x)(a_{2,1} + a_{2,0} x)^{\mu -1} x^{{a_{1,1}\over a_{2,1}}-1}dx={(-1)^{n+m}a_{2,1}^{n+m+\mu-1}(\mu)_n(\mu)_m}\left(-{a_{2,0}\over a_{2,1}}\right)^{-a_{1,1}/a_{2,1}}
{\Gamma\left({a_{1,1}\over a_{2,1}}\right)\Gamma(\mu)\over\Gamma({a_{1,1}\over a_{2,1}}+\mu)}\\
&\times \sum_{k=0}^n\sum_{j=0}^m {(-n)_k(n-1+{a_{1,0}\over a_{2,0}})_k(-m)_j(m-1+{a_{1,0}\over a_{2,0}})_j\over (\mu)_j~k!~j!}
{(k+\mu)_j\over ({a_{1,1}\over a_{2,1}}+\mu+k)_j ({a_{1,1}\over a_{2,1}}+\mu)_k}\\
&={(-1)^{n+m}a_{2,1}^{n+m+\mu-1}(\mu)_n(\mu)_m}\left(-{a_{2,0}\over a_{2,1}}\right)^{-a_{1,1}/a_{2,1}}{\Gamma\left({a_{1,1}\over a_{2,1}}\right)\Gamma(\mu)\over \Gamma({a_{1,0}\over a_{2,0}})}
\sum_{k=0}^n{(-n)_k(n-1+{a_{1,0}\over a_{2,0}})_k\over ({a_{1,1}\over a_{2,1}}+\mu)_kk!}
\\
&\times {}_3F_2(-m,m-1+{a_{1,0}\over a_{2,0}},\mu+k;{a_{1,1}\over a_{2,1}}+\mu+k,\mu;1).
\end{align*}
Using the identity
\begin{equation}\label{eq38}
{}_3F_2(-n,a,b;c,1+a+b-c-n;1)={(c-a)_n(c-b)_n\over (c)_n(c-a-b)_n},
\end{equation}
we  finally obtain
\begin{align*}
\int\limits_{0}^{{-{a_{2,1}\over a_{2,0}}}}
&y_n(x)y_m(x)(a_{2,1} + a_{2,0} x)^{\mu -1} x^{{a_{1,1}\over a_{2,1}}-1}dx={(-1)^{n+m}a_{2,1}^{n+m+\mu-1}(\mu)_n}\left(-{a_{2,0}\over a_{2,1}}\right)^{-{a_{1,1}\over a_{2,1}}}{\Gamma\left({a_{1,1}\over a_{2,1}}\right)({a_{1,1}\over a_{2,1}})_m\over ({a_{1,1}\over a_{2,1}}+\mu)_m}{\Gamma(\mu)\over \Gamma({a_{1,0}\over a_{2,0}})}\\
&\times\sum_{k=0}^n{(-n)_k(n-1+{a_{1,0}\over a_{2,0}})_k\over k!}{(-k)_m\over ({a_{1,0}\over a_{2,0}}+m)_k}\\
&=n!~~(-a_{2,0})^{-{a_{1,1}\over a_{2,1}}}~~a_{2,1}^{2n+{a_{1,0}\over a_{2,0}}-1} ~~
{\Gamma(n+{a_{1,0}\over a_{2,0}}-{a_{1,1}\over a_{2,1}})\Gamma\left(n+{a_{1,1}\over a_{2,1}}\right)\over \Gamma(n+{a_{1,0}\over a_{2,0}})}~~{(a_{1,0}+(n-1)a_{2,0})\over (a_{1,0}+(2n-1)a_{2,0})}~~\delta_{nm},
\end{align*}
where again $\delta_{nm}=0$ if $n\neq m$ and $\delta_{nm}=1$ if $n=m$.
\section*{Case III: $a_{2,1}=0$}
\vskip0.1true in
\noindent In this case the differential equation (\ref{eq3}) reads
\begin{equation}\label{eq39}
(a_{2,0}x^2+a_{2,2})~y^{\prime \prime}+(a_{1,0}x+a_{1,1})~y'-n((n-1)a_{2,0}+a_{1,0})~y=0,\quad n=0,1,2,\dots
\end{equation}
The recurrence relation to generate its polynomial solution is
\begin{align}\label{eq40}
y_{n+2}&=\left({((2n+1)a_{2,0}+a_{1,0})(2(n+1)a_{2,0}+a_{1,0})\over na_{2,0}+a_{1,0}}x+{a_{1,1}((2n+1)a_{2,0}+a_{1,0})(-2a_{2,0}+a_{1,0})\over (na_{2,0}+a_{1,0})(2na_{2,0}+a_{1,0})}\right)~y_{n+1}\notag\\
&
+{(n+1)(2(n+1)a_{2,0}+a_{1,0})(4a_{2,2}a_{2,0}^2n^2+4a_{2,0}a_{1,0}a_{2,2}n+a_{1,0}^2a_{2,2}+a_{2,0}a_{1,1}^2)\over (na_{2,0}+a_{1,0})(2na_{2,0}+a_{1,0})}~y_n
\end{align}
to be initiated with $y_0=1$ and $y_1=a_{1,0}x+a_{1,1}.$ In terms of the hypergeometric functions, the polynomial solutions can be written, for 
$a_{2,0}a_{1,1}-a_{1,0}\sqrt{-a_{2,0}a_{2,2}}\neq 0$, as
\begin{align}\label{eq41}
y_n\left(\begin{array}{lll}
a_{2,0} & 0 & a_{2,2} \\
a_{1,0} & a_{1,1}& ~ \\
\end{array}\bigg|x\right)&={(-2\sqrt{- a_{2,0}a_{2,2}})^n
\left({a_{2,0}a_{1,1}-a_{1,0}\sqrt{-a_{2,0}a_{2,2}}\over -2a_{2,0}\sqrt{-a_{2,0}a_{2,2}}}\right)_n}\notag\\
&\times
{}_2F_1
\left(\begin{array}{ll}
-n & n-1+{a_{1,0}\over a_{2,0}} \\
{a_{2,0}a_{1,1}-a_{1,0}\sqrt{-a_{2,0}a_{2,2}}\over -2a_{2,0}\sqrt{-a_{2,0}a_{2,2}}}& ~ \\
\end{array}\bigg|{a_{2,0}x+\sqrt{-a_{2,0}a_{2,2}}\over 2\sqrt{-a_{2,0}a_{2,2}}}\right)
\end{align}
while for $a_{2,0}a_{1,1}-a_{1,0}\sqrt{-a_{2,0}a_{2,2}}= 0$ as
\begin{align}\label{eq42}
y_n
\left(\begin{array}{lll}
a_{2,0} & 0 & a_{2,2} \\
a_{1,0} & {a_{1,0}\sqrt{-{a_{2,2}\over a_{2,0}}}}& ~ \\
\end{array}\bigg|x\right)
&=
\left\{ \begin{array}{ll}
 1 &\mbox{ if $n=0$} \\ 
 -n!\left(-2a_{2,0}\sqrt{- {a_{2,2}\over a_{2,0}}}\right)^n
\left(n-1+{a_{1,0}\over a_{2,0}}\right)
\left({1\over 2}
\sqrt{-{a_{2,0}\over a_{2,2}}}x+{1\over 2}\right)\\ \quad\quad\times {}_2F_1\left(1-n,n-1+{a_{1,0}\over a_{2,0}};2;{1\over 2}
\sqrt{-{a_{2,0}\over a_{2,2}}}x+{1\over 2}\right)&\mbox{if $n\geq 1$}
       \end{array} \right.\end{align}
The first few numerical coefficients of these polynomials are
\begin{equation}\label{eq43}
\left\{ \begin{array}{l}
 y_0=1 \\  
 y_1=a_{1,0}x+a_{1,1}\\ 
 y_2=(a_{1,0} + a_{2,0}) (a_{1,0} + 2 a_{2,0}) x^2+ 2 a_{1,1} (a_{1,0} + a_{2,0}) x+a_{1,1}^2 + (a_{1,0} + 2 a_{2,0}) a_{2,2} \\ 
 y_3=(a_{1,0} + 2 a_{2,0}) (a_{1,0} + 3 a_{2,0}) (a_{1,0} + 4 a_{2,0}) x^3+
 3 a_{1,1} (a_{1,0} + 2 a_{2,0}) (a_{1,0} + 3 a_{2,0}) x^2\\
 +(3 a_{1,1}^2 (a_{1,0} + 2 a_{2,0}) + 3 a_{2,2} (a_{1,0} + 2 a_{2,0}) (a_{1,0} + 4 a_{2,0})) x+a_{1,1}^3 + a_{1,1} (3 a_{1,0} + 10 a_{2,0}) a_{2,2}\\ 
 y_4=(a_{1,0} + 3 a_{2,0}) (a_{1,0} + 4 a_{2,0}) (a_{1,0} + 5 a_{2,0}) (a_{1,0} + 6 a_{2,0}) x^4+
 4 a_{1,1} (a_{1,0} + 3 a_{2,0}) (a_{1,0} + 4 a_{2,0}) (a_{1,0} + 5 a_{2,0}) x^3 \\
+(6 a_{1,1}^2 (a_{1,0} + 3 a_{2,0}) (a_{1,0} + 4 a_{2,0}) + 
   6 (a_{1,0} + 3 a_{2,0}) (a_{1,0} + 4 a_{2,0}) (a_{1,0} + 6 a_{2,0}) a_{2,2}) x^2 \\
 +(4 a_{1,1}^3 (a_{1,0} + 3 a_{2,0}) + 4 a_{1,1} a_{2,2}(a_{1,0} + 3 a_{2,0}) (3 a_{1,0} + 16 a_{2,0}) ) x\\
 +a_{1,1}^4 + 2 a_{1,1}^2 (3 a_{1,0} + 14 a_{2,0}) a_{2,2} + 
 3 (a_{1,0} + 4 a_{2,0}) (a_{1,0} + 6 a_{2,0}) a_{2,2}^2
 \\ 
  \dots
             \end{array} \right.
\end{equation}
Again, the numerical coefficients in the case  of $a_{2,0}a_{1,1}-a_{1,0}\sqrt{-a_{2,0}a_{2,2}}= 0$  are obtained by replacing $a_{1,1}$ with ${a_{1,0}\sqrt{-{a_{2,2}/a_{2,0}}}}$. The weight function and orthogonality condition depend on the value of the coefficients $a_{2,0}$ and $a_{2,2}$: We note if $ a_{2,0}>0$ and $a_{2,2}<0$, the differential equation (\ref{eq39}) can be written as
\begin{equation*}
a_{2,0}\left(\sqrt{-{a_{2,2}\over a_{2,0}}}-x\right)\left(x+\sqrt{-{a_{2,2}\over a_{2,0}}}\right)~y^{\prime \prime}-(a_{1,0}x+a_{1,1})~y'+n((n-1)a_{2,0}+a_{1,0})~y=0,\quad n=0,1,2,\dots
\end{equation*}
where $x\in\left(-\sqrt{-{a_{2,2}\over a_{2,0}}},\sqrt{-{a_{2,2}\over a_{2,0}}}\right),$ 
while for $ a_{2,0}<0$ and $a_{2,2}>0$, the differential equation (\ref{eq39}) reads
\begin{equation*}
a_{2,2}\left(1-\sqrt{-{a_{2,0}\over a_{2,2}}}x\right)\left(1+\sqrt{-{a_{2,0}\over a_{2,2}}}x\right)~y^{\prime \prime}+(a_{1,0}x+a_{1,1})~y'-n((n-1)a_{2,0}+a_{1,0})~y=0,\quad n=0,1,2,\dots
\end{equation*}
where again  $x\in\left(-\sqrt{-{a_{2,2}\over a_{2,0}}},\sqrt{-{a_{2,2}\over a_{2,0}}}\right),$ finally for $a_{2,0}\cdot a_{2,2}>0$, the differential equation (\ref{eq39}) reads
\begin{equation*}
\left(a_{2,0}x^2+a_{2,2}\right)~y^{\prime \prime}+(a_{1,0}x+a_{1,1})~y'-n((n-1)a_{2,0}+a_{1,0})~y=0,\quad n=0,1,2,\dots
\end{equation*}
where, in this case,  $x\in (-\infty,\infty)$. 
\vskip0.1 true in
\noindent $\bullet$ For $a_{2,0}>0,~a_{2,2}<0$, $a_{1,0}>0$ and $-\sqrt{-a_{1,0}^2a_{2,2}/a_{2,0}}<a_{1,1}<\sqrt{-a_{1,0}^2a_{2,2}/a_{2,0}}$ and
$x\in\left(-\sqrt{-{a_{2,2}\over a_{2,0}}},\sqrt{-{a_{2,2}\over a_{2,0}}}\right)$, Pearson's equation (\ref{eq20}) then yields 
\begin{align}\label{eq44}
W_3^1\left(\begin{array}{lll}
a_{2,0} & 0 & a_{2,2} \\
a_{1,0} & a_{1,1}& ~ \\
\end{array}\bigg|x\right)
&={1\over a_{2,0}}\left(\sqrt{-{a_{2,2}\over a_{2,0}}}-x\right)^{{1\over 2a_{2,0}}{\left(a_{1,0}+a_{1,1}\sqrt{-{a_{2,0}\over a_{2,2}}}\right)}-1}\left(\sqrt{-{a_{2,2}\over a_{2,0}}}+x\right)^{{1\over 2a_{2,0}}{\left(a_{1,0}-a_{1,1}\sqrt{-{a_{2,0}\over a_{2,2}}}\right)}-1}
\end{align}
 The orthogonality condition reads in this case 
\begin{align}\label{eq45}
&\int\limits_{-\sqrt{-{a_{2,2}\over a_{2,0}}}}^{\sqrt{-{a_{2,2}\over a_{2,0}}}} y_n(x)y_m(x)W_3^1(x)dx\notag\\
&= \left\{ \begin{array}{ll}
 0 &\mbox{\quad\quad if $m\neq n$}, \\ 
~{
2^{2n+{a_{1,0}\over a_{2,0}}-1}(-a_{2,0}a_{2,2})^{n}
{\Gamma\left(n+{a_{1,0}\over 2a_{2,0}}+{a_{1,1}\over 2a_{2,0}}
\sqrt{-{a_{2,0}\over a_{2,2}}}\right)}
\Gamma\left(n+{a_{1,0}\over 2a_{2,0}}-{a_{1,1}\over 2a_{2,0}}
\sqrt{-{a_{2,0}\over a_{2,2}}}\right)\over a_{2,0}\Gamma(n+{a_{1,0}\over a_{2,0}})\left(\sqrt{-{a_{2,2}\over a_{2,0}}}\right)^{-{a_{1,0}\over a_{2,0}}+1}}
{(a_{1,0}+(n-1)a_{2,0})\over (a_{1,0}+(2n-1)a_{2,0})}\cdot n!,
 &\mbox{\quad\quad if $m=n$.}
       \end{array} \right.
\end{align}  
which we now prove. Using the series representation of the Gauss hypergeometric function we obtain
  \begin{align*}
\int\limits_{-\sqrt{-{a_{2,2}\over a_{2,0}}}}^{\sqrt{-{a_{2,2}\over a_{2,0}}}} y_n(x)y_m(x)W_3^1(x)dx&={1\over a_{2,0}}{(-2\sqrt{- a_{2,0}a_{2,2}})^{n+m} 
\left(\mu\right)_n}{ 
\left(\mu\right)_m}\\
&\times\sum_{k=0}^n\sum_{j=0}^m {(-n)_k(n-1+{a_{1,0}\over a_{2,0}})_k\over (\mu)_k~k!}{(-m)_j(m-1+{a_{1,0}\over a_{2,0}})_j\over (\mu)_j~j!}\left({1\over 2\sqrt{-a_{2,0}a_{2,2}}}\right)^{k+j}\times I_{nm}
\end{align*}
where the integral $I_{nm}$ reads
\begin{align*}
I_{nm}
&=a_{2,0}^{k+j}\int\limits_{-\sqrt{-{a_{2,2}\over a_{2,0}}}}^{\sqrt{-{a_{2,2}\over a_{2,0}}}}
\left(\sqrt{-{a_{2,2}\over a_{2,0}}}-x\right)^{{1\over 2a_{2,0}}{\left(a_{1,0}+a_{1,1}\sqrt{-{a_{2,0}\over a_{2,2}}}\right)}-1}
\left(\sqrt{-{a_{2,2}\over a_{2,0}}}+x\right)^{k+j+{1\over 2a_{2,0}}{\left(a_{1,0}-a_{1,1}\sqrt{-{a_{2,0}\over a_{2,2}}}\right)}-1}dx\\
&=a_{2,0}^{k+j}
\left(\sqrt{-{a_{2,2}\over a_{2,0}}}\right)^{k+j+{a_{1,0}\over a_{2,0}}-1}\times \int_{-1}^1(1-\tau)^{\alpha-1}(1+\tau)^{\beta-1}d\tau,\quad 
\end{align*}
where we have used $x=\sqrt{-{a_{2,2}/a_{2,0}}}\tau$ and, for simplicity, 
$\alpha={{1\over 2a_{2,0}}{\left(a_{1,0}+a_{1,1}\sqrt{-{a_{2,0}\over a_{2,2}}}\right)}},\quad \beta={k+j+{1\over 2a_{2,0}}{\left(a_{1,0}-a_{1,1}\sqrt{-{a_{2,0}\over a_{2,2}}}\right)}}.
$
With $\tau=2\nu-1$, we can compute the integral in terms of the Beta function to yield
\begin{align*}
I_{nm}&=a_{2,0}^{k+j}
\left(\sqrt{-{a_{2,2}\over a_{2,0}}}\right)^{k+j+{a_{1,0}\over a_{2,0}}-1}
2^{{a_{1,0}\over a_{2,0}}+k+j-1}\Gamma\left({{1\over 2a_{2,0}}{\left(a_{1,0}+a_{1,1}\sqrt{-{a_{2,0}\over a_{2,2}}}\right)}}\right)\\
&\times {
\left({k+{1\over 2a_{2,0}}{\left(a_{1,0}-a_{1,1}\sqrt{-{a_{2,0}\over a_{2,2}}}\right)}}\right)_j\left({{1\over 2a_{2,0}}{\left(a_{1,0}-a_{1,1}\sqrt{-{a_{2,0}\over a_{2,2}}}\right)}}\right)_k\Gamma\left({{1\over 2a_{2,0}}{\left(a_{1,0}-a_{1,1}\sqrt{-{a_{2,0}\over a_{2,2}}}\right)}}\right)\over
({{a_{1,0}\over a_{2,0}}+k})_j({a_{1,0}\over a_{2,0}})_k\Gamma({a_{1,0}\over a_{2,0}})}.
\end{align*}
Thus
\begin{align*}
&\int\limits_{-\sqrt{-{a_{2,2}\over a_{2,0}}}}^{\sqrt{-{a_{2,2}\over a_{2,0}}}} y_n(x)y_m(x)W_3^1(x)dx\\
&={(-1)^{n+m}2^{n+m+{a_{1,0}\over a_{2,0}}-1}(\sqrt{- a_{2,0}a_{2,2}})^{n+m}{ 
\left(\mu\right)_n}{\left(\mu\right)_m}\Gamma\left({{1\over 2a_{2,0}}{\left(a_{1,0}+a_{1,1}\sqrt{-{a_{2,0}\over a_{2,2}}}\right)}}\right)\Gamma\left({{1\over 2a_{2,0}}{\left(a_{1,0}-a_{1,1}\sqrt{-{a_{2,0}\over a_{2,2}}}\right)}}\right)\over a_{2,0}\Gamma({a_{1,0}\over a_{2,0}})\left(\sqrt{-{a_{2,2}\over a_{2,0}}}\right)^{-{a_{1,0}\over a_{2,0}}+1}}
\\
&\times
\sum_{k=0}^n  {(-n)_k(n-1+{a_{1,0}\over a_{2,0}})_k
\left({{a_{1,0}-a_{1,1}\sqrt{-{a_{2,0}\over a_{2,2}}}}\over 2a_{2,0}}\right)_k\over ({a_{1,0}\over a_{2,0}})_k(\mu)_k~k!}
{}_3F_2(-m,m-1+{a_{1,0}\over a_{2,0}},
{k+{{a_{1,0}-a_{1,1}\sqrt{-{a_{2,0}\over a_{2,2}}}}\over 2a_{2,0}}};
\mu,{a_{1,0}\over a_{2,0}}+k;1)
\end{align*}
For the summation part, using $\mu=\left(a_{1,0}-a_{1,1}\sqrt{-{a_{2,0}\over a_{2,2}}}\right)/(2a_{2,0})$ and, since
\begin{align*}
{}_3F_2(-m,m-1+{a_{1,0}\over a_{2,0}},
{k+{1\over 2a_{2,0}}\bigg({{a_{1,0}-a_{1,1}\sqrt{-{a_{2,0}\over a_{2,2}}}}}}\bigg);&
{1\over 2a_{2,0}}\left(a_{1,0}-a_{1,1}\sqrt{-{a_{2,0}\over a_{2,2}}}\right),{a_{1,0}\over a_{2,0}}+k;1)\\
&={\left(-{a_{1,0}\over 2a_{2,0}}
-{a_{1,1}\over a_{2,0}} \sqrt{-{a_{2,0}\over a_{2,2}}}-m+1\right)_m\left(-k\right)_m 
\over \left({1\over 2a_{2,0}}\left(a_{1,0}-a_{1,1}\sqrt{-{a_{2,0}\over a_{2,2}}}\right)\right)_m\left(-m+1-{a_{1,0}\over a_{2,0}}-k \right)_m },
\end{align*}
and, further, since
\begin{align*}
&
\sum_{k=0}^n{(-n)_k(n-1+{a_{1,0}\over a_{2,0}})_k
\over
~k!}
{\left(-k\right)_m 
\over (m+{a_{1,0}\over a_{2,0}})_k }=
\left\{ \begin{array}{ll}
 0 &\mbox{ if $n\neq m$} \\
 {(a_{1,0}+(n-1)a_{2,0}\over a_{1,0}+(2n-1)a_{2,0}}\cdot n! &\mbox{ if $n=m$},
       \end{array} \right.
\end{align*}
we finally obtain
\begin{align*}
&\int\limits_{-\sqrt{-{a_{2,2}\over a_{2,0}}}}^{\sqrt{-{a_{2,2}\over a_{2,0}}}} y_n(x)y_m(x)W_3^1(x)dx\\
&={
2^{2n+{a_{1,0}\over a_{2,0}}-1}(-a_{2,0}a_{2,2})^{n}
{\Gamma\left(n+{a_{1,0}\over 2a_{2,0}}+{a_{1,1}\over 2a_{2,0}}
\sqrt{-{a_{2,0}\over a_{2,2}}}\right)}
\Gamma\left(n+{a_{1,0}\over 2a_{2,0}}-{a_{1,1}\over 2a_{2,0}}
\sqrt{-{a_{2,0}\over a_{2,2}}}\right)\over a_{2,0}\Gamma(n+{a_{1,0}\over a_{2,0}})\left(\sqrt{-{a_{2,2}\over a_{2,0}}}\right)^{-{a_{1,0}\over a_{2,0}}+1}}
{(a_{1,0}+(n-1)a_{2,0})\over (a_{1,0}+(2n-1)a_{2,0})}\cdot n! \cdot \delta_{nm}
\end{align*}

\vskip0.1 true in 
\noindent $\bullet$ For $a_{2,0}<0,~a_{2,2}>0$, $a_{1,0}>0$, $-\sqrt{-a_{1,0}^2a_{2,2}/a_{2,0}}<a_{1,1}<\sqrt{-a_{1,0}^2a_{2,2}/a_{2,0}}$ and 
$x\in\left(-\sqrt{-{a_{2,2}\over a_{2,0}}},\sqrt{-{a_{2,2}\over a_{2,0}}}\right),$ the weight function can be written, using equation (\ref{eq20}), as
\begin{align}\label{eq46}
W_3^2\left(\begin{array}{lll}
a_{2,0} & 0 & a_{2,2} \\
a_{1,0} & a_{1,1}& ~ \\
\end{array}\bigg|x\right)&={1\over a_{2,2}}\left(1-\sqrt{-{a_{2,0}\over a_{2,2}}}x\right)^{-{1\over 2}{\sqrt{-{a_{2,2}\over a_{2,0}}}}{\left(a_{1,1}-a_{1,0}\sqrt{-{a_{2,2}\over a_{2,0}}}\right)}-1}\left(1+\sqrt{-{a_{2,0}\over a_{2,2}}}x\right)^{{1\over 2}{\sqrt{-{a_{2,2}\over a_{2,0}}}}{\left(a_{1,1}+a_{1,0}\sqrt{-{a_{2,2}\over a_{2,0}}}\right)}-1}
\end{align} 
and the orthogonality condition can be obtained, using the series representation for Gauss hypergeometric function in (\ref{eq41}), as follows
\begin{align*}
\int\limits_{-\sqrt{-{a_{2,2}\over a_{2,0}}}}^{\sqrt{-{a_{2,2}\over a_{2,0}}}} y_n(x)y_m(x)W_3^2(x)dx&={(-2\sqrt{- a_{2,0}a_{2,2}})^{n+m}\over a_{2,2}}{ 
\left(\mu\right)_n}{ 
\left(\mu\right)_m}\\
&\times
\sum_{k=0}^n\sum_{j=0}^m {(-n)_k(n-1+{a_{1,0}\over a_{2,0}})_k\over (\mu)_k~k!}{(-m)_j(m-1+{a_{1,0}\over a_{2,0}})_j\over (\mu)_j~j!}\left({1\over 2}\right)^{k+j}\times I_{nm}
\end{align*}
where 
\begin{align*}
I_{nm}
&=\sqrt{-{a_{2,2}\over a_{2,0}}}\int\limits_{-1}^{1}
{\left(1+\tau\right)^{k+j+{a_{1,0}-a_{1,1}\sqrt{-{a_{2,0}\over a_{2,2}}}\over {2a_{2,0}}}-1}}
{\left(1-\tau\right)^{{a_{1,0}+a_{1,1}\sqrt{-{a_{2,0}\over a_{2,2}}}\over {2a_{2,0}}}-1}}d\tau
\end{align*}
using $\tau=\sqrt{-{a_{2,0}/a_{2,2}}}x.$
With $\tau=2\nu-1,$ the integral can be evaluated in terms of Beta function to yield
\begin{align*}
I_{nm}
&=2^{k+j+{a_{1,0}\over a_{2,0}}-1}
\sqrt{-{a_{2,2}\over a_{2,0}}}
{\left({k+{a_{1,0}-a_{1,1}\sqrt{-{a_{2,0}\over a_{2,2}}}\over {2a_{2,0}}}}\right)_j\left({{a_{1,0}-a_{1,1}\sqrt{-{a_{2,0}\over a_{2,2}}}\over {2a_{2,0}}}}\right)_k\Gamma\left({{a_{1,0}-a_{1,1}\sqrt{-{a_{2,0}\over a_{2,2}}}\over {2a_{2,0}}}}\right)
\Gamma\left({a_{1,0}+a_{1,1}\sqrt{-{a_{2,0}\over a_{2,2}}}\over 2a_{2,0}}\right)\over 
\left(k+{a_{1,0}\over a_{2,0}}\right)_j\left({a_{1,0}\over a_{2,0}}\right)_k\Gamma\left({a_{1,0}\over a_{2,0}}\right)}\\
\end{align*}
Consequently,
\begin{align*}
&\int\limits_{-\sqrt{-{a_{2,2}\over a_{2,0}}}}^{\sqrt{-{a_{2,2}\over a_{2,0}}}} y_n(x)y_m(x)W_3^2(x)dx
=2^{{a_{1,0}\over a_{2,0}}-1}
\sqrt{-{a_{2,2}\over a_{2,0}}}{(-2\sqrt{- a_{2,0}a_{2,2}})^{n+m}\over a_{2,2}}{ 
\left(\mu\right)_n}{ 
\left(\mu\right)_m}{\Gamma\left({{a_{1,0}-a_{1,1}\sqrt{-{a_{2,0}\over a_{2,2}}}\over {2a_{2,0}}}}\right)
\Gamma\left({a_{1,0}+a_{1,1}\sqrt{-{a_{2,0}\over a_{2,2}}}\over 2a_{2,0}}\right)\over \Gamma\left({a_{1,0}\over a_{2,0}}\right)}\\
&\times
\sum_{k=0}^n{(-n)_k(n-1+{a_{1,0}\over a_{2,0}})_k\over (\mu)_k~k!}{\left({{a_{1,0}-a_{1,1}\sqrt{-{a_{2,0}\over a_{2,2}}}\over {2a_{2,0}}}}\right)_k\over \left({a_{1,0}\over a_{2,0}}\right)_k}
{}_3F_2(-m,m-1+{a_{1,0}\over a_{2,0}},{k+{a_{1,0}-a_{1,1}\sqrt{-{a_{2,0}\over a_{2,2}}}\over {2a_{2,0}}}};{a_{1,0}-a_{1,1}\sqrt{-{a_{2,0}\over a_{2,2}}}\over {2a_{2,0}}},k+{a_{1,0}\over a_{2,0}};1)
\end{align*}
again using the  identity
$${}_3F_2(-m,a,b;c,1+a+b-c-m;1)={(c-a)_m(c-b)_m\over (c)_m(c-a-b)_m}
$$
we finally have
\begin{align*}
&\int\limits_{-\sqrt{-{a_{2,2}\over a_{2,0}}}}^{\sqrt{-{a_{2,2}\over a_{2,0}}}} y_n(x)y_m(x)W_3^2(x)dx
=2^{{a_{1,0}\over a_{2,0}}-1}
\sqrt{-{a_{2,2}\over a_{2,0}}}{(-2\sqrt{- a_{2,0}a_{2,2}})^{n+m}\over a_{2,2}}\\
&\times {\Gamma\left(n+{{a_{1,0}-a_{1,1}\sqrt{-{a_{2,0}\over a_{2,2}}}\over {2a_{2,0}}}}\right)
\Gamma\left(m+{a_{1,0}+a_{1,1}\sqrt{-{a_{2,0}\over a_{2,2}}}\over 2a_{2,0}}\right)}
\times {1\over\Gamma\left(m+{a_{1,0}\over a_{2,0}}\right)}\times \left\{ \begin{array}{ll}
0 &\mbox{ if $n\neq m$,} \\
  n!{(a_{1,0}+(n-1)a_{2,0})\over(a_{1,0}+(2n-1)a_{2,0}) }&\mbox{ if $n= m$.} 
       \end{array} \right. \\
&=2^{2n+{a_{1,0}\over a_{2,0}}-1}
\sqrt{-{a_{2,2}\over a_{2,0}}}{(- a_{2,0}a_{2,2})^{n}\over a_{2,2}}{\Gamma\left(n+{{a_{1,0}-a_{1,1}\sqrt{-{a_{2,0}\over a_{2,2}}}\over {2a_{2,0}}}}\right)
\Gamma\left(n+{a_{1,0}+a_{1,1}\sqrt{-{a_{2,0}\over a_{2,2}}}\over 2a_{2,0}}\right)}
\\
&\times {n!\over\Gamma\left(n+{a_{1,0}\over a_{2,0}}\right)}{(a_{1,0}+(n-1)a_{2,0})\over(a_{1,0}+(2n-1)a_{2,0}) }~\delta_{nm}.
 \end{align*}

\noindent $\bullet$ For $ a_{2,0}\cdot a_{2,2}>0$, Pearson's equation reads 
\begin{align}\label{eq47}
W_3^3\left(\begin{array}{lll}
a_{2,0} & 0 & a_{2,2} \\
a_{1,0} & a_{1,1}& ~ \\
\end{array}\bigg|x\right)
&=\left(a_{2,0}x^2+a_{2,2}\right)^{{a_{1,0}\over 2a_{2,0}}-1}\exp\left({a_{1,1}\over a_{2,0}}\sqrt{a_{2,0}\over a_{2,2}}\arctan\left(\sqrt{a_{2,0}\over a_{2,2}}x\right)\right),
\end{align}
where $x\in (-\infty, \infty)$. The orthogonality relation reads
\begin{align*}
\int_{-\infty}^\infty y_n(x)y_m(x) W_3^3(x)dx
&={(-2\sqrt{- a_{2,0}a_{2,2}})^{n+m}(\mu)_n(\mu)_m}\\
&\times
\sum_{k=0}^n\sum_{j=0}^m {(-n)_k(-m)_j(n-1+{a_{1,0}\over a_{2,0}})_k(m-1+{a_{1,0}\over a_{2,0}})_j\over (\mu)_k(\mu)_j~k!~j!}\left({1\over 2\sqrt{-a_{2,0}a_{2,2}}}\right)^{k+j}\times I_{nm}
\end{align*}
where
\begin{align*}
I_{nm}=& \int_{-\infty}^\infty
\left({a_{2,0}x+\sqrt{-a_{2,0}a_{2,2}}}\right)^{k+j}
\left(a_{2,0}x^2+a_{2,2}\right)^{{a_{1,0}\over 2a_{2,0}}-1}\exp\left({a_{1,1}\over a_{2,0}}\sqrt{a_{2,0}\over a_{2,2}}\arctan\left(\sqrt{a_{2,0}\over a_{2,2}}x\right)\right)dx
\end{align*}
Using the identity
$\arctan(z)={i\over 2}\left(\ln(1-iz)-\ln(1+iz)\right)$ where $i=\sqrt{-1}$, the integral reads
\begin{align*}
I_{nm}
&=i^{k+j} (a_{2,0})^{k+j+{a_{1,0}\over 2a_{2,0}}-1}
\int_{-\infty}^\infty
\left({\sqrt{a_{2,2}\over a_{2,0}}+ix}\right)^{{a_{1,0}\over 2a_{2,0}}-{i\over 2}{a_{1,1}\over a_{2,0}}\sqrt{a_{2,0}\over a_{2,2}}-1}
\left({\sqrt{{a_{2,2}\over a_{2,0}}}-ix}\right)^{k+j+{a_{1,0}\over 2a_{2,0}}+{{i\over 2}{a_{1,1}\over a_{2,0}}\sqrt{a_{2,0}\over a_{2,2}}}-1}dx
\end{align*}
further using the identity
$$\int_{-\infty}^\infty {dt\over (a+it)^c(b-it)^d}=2\pi{\Gamma(c+d-1)\over \Gamma(c)\Gamma(d)}(a+b)^{1-(c+d)}, Re(c+d)>1,Re~ a>0,Re~b>0
$$
with
$a=b=\sqrt{{a_{2,2}\over a_{2,0}}}>0$, $c=-{{a_{1,0}\over 2a_{2,0}}+{i\over 2}{a_{1,1}\over a_{2,0}}\sqrt{a_{2,0}\over a_{2,2}}+1}
$ and $d={-k-j-{a_{1,0}\over 2a_{2,0}}-{{i\over 2}{a_{1,1}\over a_{2,0}}\sqrt{a_{2,0}\over a_{2,2}}}+1}$, we have for $1-{a_{1,0}\over a_{2,0}}>k+j$,
after some simplifications using Pochhammer's identities, 
\begin{align*}
I_{nm}
&={\pi i^{k+j}2^{{a_{1,0}\over a_{2,0}}+k+j}a_{2,0}^{{a_{1,0}\over 2a_{2,0}}+k+j-1}
\left(\sqrt{{a_{2,2}\over a_{2,0}}}\right)^{{a_{1,0}\over a_{2,0}}+k+j-1}\Gamma(-{a_{1,0}\over a_{2,0}}+1)\over \Gamma(-{{a_{1,0}\over 2a_{2,0}}+{i\over 2}{a_{1,1}\over a_{2,0}}\sqrt{a_{2,0}\over a_{2,2}}+1})\Gamma({-{a_{1,0}\over 2a_{2,0}}-{{i\over 2}{a_{1,1}\over a_{2,0}}\sqrt{a_{2,0}\over a_{2,2}}}+1})}
{(k+{{a_{1,0}\over 2a_{2,0}}+{{i\over 2}{a_{1,1}\over a_{2,0}}\sqrt{a_{2,0}\over a_{2,2}}}})_{j}({{a_{1,0}\over 2a_{2,0}}+{{i\over 2}{a_{1,1}\over a_{2,0}}\sqrt{a_{2,0}\over a_{2,2}}}})_{k} \over 
(k+{a_{1,0}\over a_{2,0}})_{j}({a_{1,0}\over a_{2,0}})_{k} }\\\end{align*}
Thus
\begin{align*}
\int_{-\infty}^\infty y_n(x)y_m(x) W_3^3(x)dx&={{(-2\sqrt{- a_{2,0}a_{2,2}})^{n+m}(\mu)_n(\mu)_m}2^{{a_{1,0}\over a_{2,0}}}
\pi(a_{2,0})^{{a_{1,0}\over 2a_{2,0}}-1}\left(\sqrt{{a_{2,2}\over a_{2,0}}}\right)^{{a_{1,0}\over a_{2,0}}-1}\Gamma(-{a_{1,0}\over a_{2,0}}+1)\over \Gamma(-{{a_{1,0}\over 2a_{2,0}}+{i\over 2}{a_{1,1}\over a_{2,0}}\sqrt{a_{2,0}\over a_{2,2}}+1})\Gamma({-{a_{1,0}\over 2a_{2,0}}-{{i\over 2}{a_{1,1}\over a_{2,0}}\sqrt{a_{2,0}\over a_{2,2}}}+1})}\\&\times\sum_{k=0}^n {(-n)_k(n-1+{a_{1,0}\over a_{2,0}})_k\over~({a_{1,0}\over a_{2,0}})_{k}~k!}
{}_3F_2(-m,m-1+{a_{1,0}\over a_{2,0}},k+{{a_{1,0}\over 2a_{2,0}}+{{i\over 2}{a_{1,1}\over a_{2,0}}\sqrt{a_{2,0}\over a_{2,2}}}};\mu,k+{a_{1,0}\over a_{2,0}};1)
\end{align*}
where $\mu=({{a_{1,0}}+i{a_{1,1}\sqrt{a_{2,0}/a_{2,2}}}})/(2a_{2,0})
$. Again using
${}_3F_2(-m,a,b;c,1+a+b-c-m;1)={(c-a)_m(c-b)_m\over [(c)_m(c-a-b)_m]},
$
we have
\begin{align*}
{}_3F_2(-m,m-1+{a_{1,0}\over a_{2,0}},k+{{a_{1,0}\over 2a_{2,0}}+{{i\over 2}{a_{1,1}\over a_{2,0}}\sqrt{a_{2,0}\over a_{2,2}}}};{{a_{1,0}\over 2a_{2,0}}+{{i\over 2}{a_{1,1}\over a_{2,0}}\sqrt{a_{2,0}\over a_{2,2}}}},k+{a_{1,0}\over a_{2,0}};1)={({{a_{1,0}\over 2a_{2,0}}-{{i\over 2}{a_{1,1}\over a_{2,0}}\sqrt{a_{2,0}\over a_{2,2}}}})_m(-k)_m\over ({{a_{1,0}\over 2a_{2,0}}+{{i\over 2}{a_{1,1}\over a_{2,0}}\sqrt{a_{2,0}\over a_{2,2}}}})_m
({a_{1,0}\over a_{2,0}}+k)_m}
\end{align*}
However, since
$\left({a_{1,0}\over a_{2,0}}+k\right)_m={(m+a_{1,0}/a_{2,0})_k({a_{1,0}/a_{2,0}})_m/ ({a_{1,0}/a_{2,0}})_k}$,
we finally have
\begin{align*}
\int_{-\infty}^\infty y_n(x)y_m(x) W_3^3(x)dx
&={{2^{2n+{{a_{1,0}\over a_{2,0}}}}({- a_{2,0}a_{2,2}})^{n} \left({{a_{1,0}\over 2a_{2,0}}+{{a_{1,1}
\over 2a_{2,0}}\sqrt{-{a_{2,0}\over a_{2,2}}}}}\right)_n}
\pi(a_{2,0})^{{a_{1,0}\over 2a_{2,0}}-1}\left({{a_{2,2}\over a_{2,0}}}\right)^{{a_{1,0}\over 2a_{2,0}}-{1\over 2}}\Gamma(-{a_{1,0}\over a_{2,0}}+1)\over \Gamma(-{{a_{1,0}\over 2a_{2,0}}+{a_{1,1}\over 2a_{2,0}}\sqrt{-{a_{2,0}\over a_{2,2}}}+1})\Gamma({-{a_{1,0}\over 2a_{2,0}}-{{a_{1,1}\over 2a_{2,0}}\sqrt{-{a_{2,0}\over a_{2,2}}}}+1})}\\
&\times
{\left({{a_{1,0}\over 2a_{2,0}}-{{a_{1,1}\over 2a_{2,0}}\sqrt{-{a_{2,0}\over a_{2,2}}}}}\right)_n\over 
({a_{1,0}\over a_{2,0}})_n}{(a_{1,0}+(n-1)a_{2,0})
\over (a_{1,0}+(2n-1)a_{2,0})}~n!~\delta_{nm},
\end{align*}
valid for $a_{1,0}/a_{2,0}<1-2n, ~n=0,1,2,\dots$
Note: we may use the identity
$(a+ib)_n(a-ib)_n=\prod_{k=0}^{n-1}((a+k)^2+b^2)$ for $i=\sqrt{-1}$ to simplify this equation further.
 \vskip0.1true in
\section*{Case IV: $a_{2,0}=a_{2,1}=0$}
\vskip0.1true in
\noindent In this case, equation (\ref{eq3}) reads
\begin{align}\label{eq48}
a_{2,2}~y''+(a_{1,0}x+a_{1,1})~y'-n~a_{1,0}~y=0,\quad n=0,1,2,\dots
\end{align}
and the recurrence relation that generates the polynomial solutions is 
\begin{equation}\label{eq49}
\left\{ \begin{array}{l}
 y_{n+2}=\left(a_{1,0}x+a_{1,1}\right)~y_{n+1}+
(n+1)a_{1,0}a_{2,2}~y_n, \\  
  y_0=1,\quad y_1=a_{1,0}x+a_{1,1}.
       \end{array} \right.
\end{equation}
These polynomial solutions are be expressed in terms of hypergeometric functions as 
\begin{align}\label{eq50}
y_n\left(\begin{array}{lll}
0 & 0 & a_{2,2} \\ 
a_{1,0} & a_{1,1}& ~ \\ 
\end{array}\bigg|x\right)&=(a_{1,0}x+a_{1,1})^n{}_2F_0\left(-{n\over 2},{-(n-1)\over 2};~-~
;{2a_{2,2}a_{1,0}\over (a_{1,0}x+a_{1,1})^2 }\right).
\end{align}
The first few numerical coefficients of the polynomial solutions are 
\begin{equation}\label{eq53}
\left\{ \begin{array}{l}
 y_0=1 \\  
 y_1=a_{1,0}x+a_{1,1}\\ 
 y_2=a_{1,0}^2x^2+2a_{1,0}a_{1,1}x+a_{1,1}^2+a_{1,0}a_{2,2}\\
 y_3=a_{1,0}^3x^3+3a_{1,0}^2a_{1,1}x^2+3a_{1,0}(a_{1,1}^2+a_{1,0}a_{2,2})x+a_{1,1}^3+3a_{1,1}a_{1,0}a_{2,2}\\ 
 y_4=a_{1,0}^4x^4+4a_{1,0}^3a_{1,1}x^3
 +6a_{1,0}^2(a_{1,1}^2+a_{1,0}a_{2,2})x^2+4a_{1,0}a_{1,1}(a_{1,1}^2+3a_{1,0}a_{2,2})x \\
+6a_{1,0}a_{2,2}a_{1,1}^2+3a_{1,0}^2a_{2,2}^2+a_{1,1}^4\\
\dots
             \end{array} \right.
\end{equation}
\vskip0.1true in
\noindent For the weight function and orthogonality relations we have for $a_{2,2}>0$, $a_{1,0}<0$ and $x\in (-\infty,\infty)$, by means of the Pearson's equation, the weight function reads
\begin{align}\label{eq52}
W_4\left(\begin{array}{lll}
0 &0 & a_{2,2} \\
a_{1,0} & a_{1,1}& ~ \\
\end{array}\bigg|x\right)
&=
{1\over a_{2,2}}\exp\left({x(a_{1,0}x+2a_{1,1})\over 2a_{2,2}}\right)={1\over a_{2,2}}
\exp\left(-{a_{1,1}^2\over 2a_{2,2}a_{1,0}}\right)\exp\left({(a_{1,0}x + a_{1,1})^2\over 2a_{1,0}a_{2,2}}\right).
\end{align}
For simplicity, we consider two separate cases:
If $n$ is an even integer, i.e. $n=2\nu, \nu=0,1,2,\dots$, the polynomial solutions are given, for $a_{1,0}<0$ and $a_{2,2}>0$, by
\begin{align}\label{eq53}
y_n\left(\begin{array}{lll}
0 & 0 & a_{2,2} \\ 
a_{1,0} & a_{1,1}& ~ \\ 
\end{array}\bigg|x\right)&=(a_{1,0}x+a_{1,1})^{2\nu}
{}_2F_0\left(-\nu,-\nu+{1\over 2};~-~
;{2a_{2,2}a_{1,0}\over (a_{1,0}x+a_{1,1})^2 }\right).
\end{align}
In this case we have for $a_{1,0}<0$ and $a_{2,2}>0$,  the orthogonality relation
\begin{align}\label{eq54}
\int\limits_{-\infty}^\infty y_{2\nu}\left(\begin{array}{lll}
0 & 0 & a_{2,2} \\
a_{1,0} & a_{1,1}& ~ \\
\end{array}\bigg|x\right)&y_{2\nu}\left(\begin{array}{lll}
0 & 0 & a_{2,2} \\
a_{1,0} & a_{1,1}& ~ \\
\end{array}\bigg|x\right)W_4(x)~dx\notag\\
&= \left\{ \begin{array}{ll}
 0 &\mbox{\quad\quad if $m\neq n$}, \\ 
 \sqrt{-{a_{1,0}^3\over a_{2,2}}}
(-{a_{1,0} a_{2,2}})^{2{\nu}}
  \exp\left(-{a_{1,1}^2\over 2a_{2,2}a_{1,0}}\right)
\sqrt{2\pi}\Gamma(2\nu+1),
 &\mbox{\quad\quad if $m=n$.}
       \end{array} \right.
\end{align}
while if $n=2\nu+1, \nu=0,1,2,\dots$ is an odd number, the polynomial solution reads
\begin{align}\label{eq55}
y_n\left(\begin{array}{lll}
0 & 0 & a_{2,2} \\ 
a_{1,0} & a_{1,1}& ~ \\ 
\end{array}\bigg|x\right)&=(a_{1,0}x+a_{1,1})^{2\nu+1}{}_2F_0\left(-\nu-{1\over 2},-\nu;~-~
;{2a_{2,2}a_{1,0}\over (a_{1,0}x+a_{1,1})^2 }\right),
\end{align}
and in this case the corresponding orthogonality relation reads
\begin{align}\label{eq56}
\int\limits_{-\infty}^\infty y_{2\nu+1}\left(\begin{array}{lll}
0 & 0 & a_{2,2} \\
a_{1,0} & a_{1,1}& ~ \\
\end{array}\bigg|x\right)&y_{2\nu+1}\left(\begin{array}{lll}
0 & 0 & a_{2,2} \\
a_{1,0} & a_{1,1}& ~ \\
\end{array}\bigg|x\right)W_4(x)~dx\notag\\
&= \left\{ \begin{array}{ll}
 0 &\mbox{\quad\quad if $m\neq n$}, \\ 
a_{1,0}^2
~(-a_{1,0}a_{2,2})^{2\nu+{1\over 2}}~\exp\left(-{a_{1,1}^2\over 2a_{2,2}a_{1,0}}\right)\sqrt{2\pi}~\Gamma(2\nu+2),
 &\mbox{\quad\quad if $m=n$.}
       \end{array} \right.
\end{align}
To prove these identities, i.e. (\ref{eq54}) and (\ref{eq56}), 
we consider the first case,  $n$ is even number, using the series representation of the hypergeometric function ${}_2F_0(-n,a;-;x)=\sum_{k=0}^n (-n)_k(a)_k x^k/k!$, we have  
\begin{align*}
\int_{-\infty}^\infty y_{2\nu}(x)y_{2\nu'}(x)W_4(x)dx&={a_{1,0}\over a_{2,2}}
\sum_{k=0}^\nu \sum_{j=0}^{\nu'}
{(-\nu)_k (-\nu+{1\over 2})_k\over k!}
{(-\nu')_j(-\nu'+{1\over 2})_j\over j!}\left(2a_{2,2}a_{1,0}\right)^{k+j}
\exp\left(-{a_{1,1}^2\over 2a_{2,2}a_{1,0}}\right)
\\
&\times \int_{-\infty}^\infty\tau^{2\nu+2\nu'-2k-2j} \exp\left({\tau^2\over 2a_{1,0}a_{2,2}}\right)dx\\
\end{align*}
where $\tau=a_{1,0}x+a_{1,1}$, the integral can be easily computed in terms of Gamma function to yield
\begin{align*}
\int_{-\infty}^\infty\tau^{2\nu+2\nu'-2k-2j} \exp\left({\tau^2\over 2a_{1,0}a_{2,2}}\right)dx&= (-2{a_{1,0} a_{2,2}})^{{1\over 2} + {\nu} + {\nu'}}
{\Gamma\left({1\over 2}+ {\nu}+ {\nu'}\right) ({1\over 2a_{1,0} a_{2,2}})^{j + k}\over 
\left(k+{1\over 2}- {\nu}- {\nu'}\right)_j\left({1\over 2}- {\nu}- {\nu'}\right)_{k}}\\
\end{align*}
where $-j-k+\nu+\nu'>-1/2$ and  $a_{1,0}<0$. Thus, we obtain after some simplification that
\begin{align*}
\int_{-\infty}^\infty y_{2\nu}(x)y_{2\nu'}(x)W_4(x)dx&={a_{1,0}\over a_{2,2}}
(-2{a_{1,0} a_{2,2}})^{{1\over 2} + {\nu} + {\nu'}}
  \Gamma\left({1\over 2}+ {\nu}+ {\nu'}\right)
  \exp\left(-{a_{1,1}^2\over 2a_{2,2}a_{1,0}}\right)
  \\
  &\times 
\sum_{k=0}^\nu
{(-\nu)_k (-\nu+{1\over 2})_k\over \left({1\over 2}- {\nu}- {\nu'}\right)_{k}k!}
{}_2F_1(-\nu',-\nu'+{1\over 2};k+{1\over 2}- {\nu}- {\nu'};1)
\end{align*}
where we have used the series representation of the Gauss hypergeometric function. Now, since ${}_2F_1(-n,a;c;1)=(c-a)_n/(c)_n$, we obtain
\begin{align*}
\int_{-\infty}^\infty y_{2\nu}(x)y_{2\nu'}(x)W_4(x)dx
&={a_{1,0}\over a_{2,2}}
(-2{a_{1,0} a_{2,2}})^{{1\over 2} + {\nu} + {\nu'}}
  \Gamma\left({1\over 2}+ {\nu}+ {\nu'}\right)
  \exp\left(-{a_{1,1}^2\over 2a_{2,2}a_{1,0}}\right)
  \\
  &\times 
\sum_{k=0}^\nu
{(-\nu)_k (-\nu+{1\over 2})_k\over \left({1\over 2}- {\nu}- {\nu'}\right)_{k}k!}
{(k- {\nu})_{\nu'}\over (k+{1\over 2}- {\nu}- {\nu'})_{\nu'}}
\end{align*}
Using the identity
$${(k- {\nu})_{\nu'}\over (k+{1\over 2}- {\nu}- {\nu'})_{\nu'} }
={(- {\nu})_{\nu'}(\nu'- {\nu})_{k}\over(- {\nu})_k }\times{ ({1\over 2}- {\nu}- {\nu'})_{k}
 \over  ({1\over 2}- {\nu}- {\nu'})_{\nu'} 
 ({1\over 2}- {\nu})_k},
$$
we obtain finally
\begin{align*}
\int_{-\infty}^\infty y_{2\nu}(x)y_{2\nu'}(x)W_4(x)dx
&={a_{1,0}\over a_{2,2}}
(-{a_{1,0} a_{2,2}})^{{1\over 2} + 2{\nu}}
  \exp\left(-{a_{1,1}^2\over 2a_{2,2}a_{1,0}}\right)
\sqrt{2\pi}\Gamma(2\nu+1)\delta_{\nu\nu'}\\
\end{align*}
where $\delta_{\nu\nu'}=0$ if $\nu\neq \nu'$ and $\delta_{\nu\nu'}=1$ if $\nu=\nu'$. 
For the case $n=2\nu+1$, $\nu=0,1,2,\dots$, we note, using the series representation of ${}_2F_0$, that
\begin{align*}
\int_{-\infty}^\infty y_{2\nu+1}(x)y_{2\nu'+1}(x)W_4(x)dx&=
{ a_{1,0}(-2a_{1,0}a_{2,2})^{{3\over 2}+\nu+\nu'}\over a_{2,2}}\exp\left(-{a_{1,1}^2\over 2a_{2,2}a_{1,0}}\right)\Gamma({3\over 2}+\nu+\nu')\\
&\times \sum_{k=0}^\nu \sum_{j=0}^{\nu'}
{(-\nu)_k (-\nu-{1\over 2})_k\over k!}
{(-\nu')_j(-\nu'-{1\over 2})_j\over j!}
{1\over (k-{1\over 2}-\nu-\nu')_{j}(-{1\over 2}-\nu-\nu')_{k}}
\end{align*}
which implies (by using the series representation of the Gauss hypergeometric function) that
\begin{align*}
\int_{-\infty}^\infty y_{2\nu+1}(x)y_{2\nu'+1}(x)W_4(x)dx&={ a_{1,0}(-2a_{1,0}a_{2,2})^{{3\over 2}+\nu+\nu'}\over a_{2,2}}\exp\left(-{a_{1,1}^2\over 2a_{2,2}a_{1,0}}\right)\Gamma({3\over 2}+\nu+\nu') \sum_{k=0}^\nu
{1\over k!}{{(-\nu)_{\nu'}(\nu'-\nu)_k} \over 
{(-{1\over 2}-\nu-\nu')_{\nu'}}}.
\end{align*}
Since
$\sum_{k=0}^\nu
{1\over k!}{{(-\nu)_{\nu'}(\nu'-\nu)_k} \over 
{(-{1\over 2}-\nu-\nu')_{\nu'}}}=0
$ if $\nu\neq \nu'$ and
$
{\nu!\over 
({3\over 2}+\nu)_{\nu}}$ if $\nu=\nu'$
we obtain finally
\begin{align*}
\int_{-\infty}^\infty y_{2\nu+1}(x)y_{2\nu'+1}(x)W_4(x)dx
&=a_{1,0}^2
~(-a_{1,0}a_{2,2})^{2\nu+{1\over 2}}~\exp\left(-{a_{1,1}^2\over 2a_{2,2}a_{1,0}}\right)\sqrt{2\pi}~\Gamma(2\nu+2)~\delta_{\nu\nu'},
\end{align*}
as noted in equation (\ref{eq56}).

\section*{Case V: $a_{2,0}=a_{2,2}=0$}
\noindent In this case, the differential equation (\ref{eq3}) reduces to
\begin{equation}\label{eq57}
a_{2,1}~x~y^{\prime \prime}+(a_{1,0}x+a_{1,1})~y'-n~a_{1,0}~y=0
\end{equation}
and the recurrence relation that generates the polynomial solutions is
\begin{align}\label{eq58}
y_{n+2}&=(a_{1,0}x+
{2(n+1)a_{2,1}+a_{1,1}})~y_{n+1}
-a_{2,1}{(n+1)(a_{2,1}n+a_{1,1})}~y_n, ~y_0(x)=1,~y_1(x)=a_{1,0}x+a_{1,1}.
\end{align}
These polynomial solutions can be expressed in terms of the confluent hypergeometric function as 
\begin{align}\label{eq59}
y_n\left(\begin{array}{lll}
0 & a_{2,1} & 0 \\
a_{1,0} & a_{1,1}& ~ \\
\end{array}\bigg|x\right)&=a_{2,1}^n\left({a_{1,1} \over a_{2,1}}\right)_n{}_1F_1\left(-n;{a_{1,1} \over a_{2,1}}
;-{a_{1,0}\over a_{2,1}}x \right),
\end{align}
which follows directly from equation (\ref{eq24}) with $a_{2,2}\rightarrow 0$. The numerical coefficient for the first few polynomials are
\begin{equation}\label{eq60}
\left\{ \begin{array}{l}
 y_0=1, \\  
 y_1=a_{1,0}x+a_{1,1},\\
 y_2=a_{1,0}^2  x^2+ 2 a_{1,0}(a_{1,1}+a_{2,1})x+a_{1,1}(a_{1,1}+a_{2,1}), \\ 
 y_3=a_{1,0}^3x^3+ 3 a_{1,0}^2(a_{1,1}+2a_{2,1}) x^2+3a_{1,0}(a_{1,1}+a_{2,1})( a_{1,1}+2a_{2,1})x + a_{1,1}(a_{1,1}+a_{2,1})(a_{1,1}+2a_{2,1}),\\ 
y_4=a_{1,0}^4x^4+4a_{1,0}^3(a_{1,1}+3a_{2,1})x^3+6a_{1,0}^2(a_{1,1}+2a_{2,1})(a_{1,1}+3a_{2,1})x^2\\
\quad~~+4a_{1,0}(a_{1,1}+3a_{2,1})(a_{1,1}+2a_{2,1})(a_{1,1}+a_{2,1})x+ a_{1,1}(a_{1,1}+a_{2,1})(a_{1,1}+2a_{2,1})(a_{1,1}+3a_{2,1}),
 \\ 
  \dots.
             \end{array} \right.
\end{equation}
For the weight function and orthogonality relation, we note, from the Pearson's equation that for ${a_{1,0}/a_{2,1}}<0$ and ${a_{1,1}/a_{2,1}}>0$, that
\begin{align}\label{eq61}
W_5\left(\begin{array}{lll}
0 & a_{2,1} & 0 \\
a_{1,0} & a_{1,1}& ~ \\
\end{array}\bigg|x\right)
&={1\over a_{2,1}}e^{{a_{1,0}\over a_{2,1}}x}x^{{a_{1,1}\over a_{2,1}}-1},
\end{align}
where $x\in(0,\infty)$ and the orthogonality relation then reads
for ${a_{1,0}/ a_{2,1}}<0$ and ${a_{1,1}/a_{2,1}}>0$,
\begin{align}\label{eq62}
&\int\limits_{0}^\infty
y_n\left(\begin{array}{lll}
0 & a_{2,1} & 0 \\
a_{1,0} & a_{1,1}& ~ \\
\end{array}\bigg|x\right)y_m\left(\begin{array}{lll}
0 & a_{2,1} & 0 \\
a_{1,0} & a_{1,1}& ~ \\
\end{array}\bigg|x\right)W_5(x)dx= \left\{ \begin{array}{ll}
 0, &\mbox{\quad if $m\neq n$}, \\
n!~a_{2,1}^{2n-1}~\left(-{a_{1,0}\over a_{2,1}}\right)^{-{a_{1,1}\over a_{2,1}}}
\Gamma(n+{a_{1,1}\over a_{2,1}}),&\mbox{\quad if $m=n$.}
       \end{array} \right.
\end{align}
This proof of this identity follows directly using the series representation of the Confluent hypergeometric function and the integral representation of the Gamma function, thus the details are omitted

\section*{Case VI: $a_{2,1}=a_{2,2}=0$}
\noindent In this final case, the differential equation (\ref{eq3}) is reduced to 
\begin{equation}\label{eq63}
a_{2,0}~x^2~y^{\prime \prime}+(a_{1,0}x+a_{1,1})~y'-n((n-1)a_{2,0}+a_{1,0})~y=0,\quad n=0,1,2,\dots
\end{equation}
and the recurrence relations for the exact solutions are
\begin{align}\label{eq66}
y_{n+2}&=\left({((2n+1)a_{2,0}+a_{1,0})(2(n+1)a_{2,0}+a_{1,0})\over na_{2,0}+a_{1,0}}x+{a_{1,1}((2n+1)a_{2,0}+a_{1,0})(-2a_{2,0}+a_{1,0})\over (na_{2,0}+a_{1,0})(2na_{2,0}+a_{1,0})}\right)~y_{n+1}\notag\\
&
+{a_{2,0}a_{1,1}^2(n+1)(2(n+1)a_{2,0}+a_{1,0})\over (na_{2,0}+a_{1,0})(2na_{2,0}+a_{1,0})}~y_n
\end{align}
to be initiated with $y_0=1,\quad y_1=a_{1,0}x+a_{1,1}.$ These polynomial solutions can be expressed in terms of the hypergeometric function as 
\begin{align}\label{eq65}
y_n\left(\begin{array}{lll}
a_{2,0} & 0 & 0 \\
a_{1,0} & a_{1,1}& ~ \\
\end{array}\bigg|x\right)&=a_{1,1}^n~ {}_2F_0\left(-n,n-1+{a_{1,0}\over a_{2,0}};~-~
;-{a_{2,0}\over a_{1,1}}x\right)
\end{align}
which follows directly from equation (\ref{eq33}) using the series representation of the Gauss hypergeometric series ${}_2F_1$ and the identity $\lim_{a_{2,1}\rightarrow 0} a_{2,1}^n(a_{1,0}/a_{2,0}-a_{1,1}/a_{2,1})_n=(-a_{1,1})^n$. The first few numerical coefficients of the polynomial solutions are
\begin{equation}\label{eq66}
\left\{ \begin{array}{l}
 y_0=1 \\
 y_1=a_{1,0}x+a_{1,1}\\
 y_2=(a_{1,0} + a_{2,0}) (a_{1,0} + 2 a_{2,0}) x^2+ 2 a_{1,1} (a_{1,0} + a_{2,0}) x+a_{1,1}^2 \\ 
 y_3=(a_{1,0} + 2 a_{2,0}) (a_{1,0} + 3 a_{2,0}) (a_{1,0} + 4 a_{2,0}) x^3+
 3 a_{1,1} (a_{1,0} + 2 a_{2,0}) (a_{1,0} + 3 a_{2,0}) x^2 +3 a_{1,1}^2 (a_{1,0} + 2 a_{2,0}) x+a_{1,1}^3 \\
 y_4=(a_{1,0} + 3 a_{2,0}) (a_{1,0} + 4 a_{2,0}) (a_{1,0} + 5 a_{2,0}) (a_{1,0} + 6 a_{2,0}) x^4+
 4 a_{1,1} (a_{1,0} + 3 a_{2,0}) (a_{1,0} + 4 a_{2,0}) (a_{1,0} + 5 a_{2,0}) x^3 \\
+6 a_{1,1}^2(a_{1,0} + 3 a_{2,0}) (a_{1,0} + 4 a_{2,0})  x^2 
 +4 a_{1,1}^3 (a_{1,0} + 3 a_{2,0}) x +a_{1,1}^4 
 \\ 
  \dots
             \end{array} \right.
\end{equation}
For the weight function and orthogonality condition, we note that from the Pearson's equation we have
\begin{align}\label{eq67}
W_6\left(\begin{array}{lll}
a_{2,0} & 0 & 0 \\
a_{1,0} & a_{1,1}& ~ \\
\end{array}\bigg|x\right)
&={1\over a_{2,0}}e^{-{a_{1,1}\over a_{2,0}x}}x^{{a_{1,0}\over a_{2,0}}-2}
\end{align}
where $x\in (0,\infty)$. Further, for ${a_{1,0}\over a_{2,0}}<1-2n$ and ${a_{1,1}/a_{2,0}}>0$, 
\begin{align}\label{eq68}
\int\limits_{0}^\infty
y_n(x)y_m(x)W_6(x)dx
= \left\{ \begin{array}{ll}
 0, &\mbox{\quad if $m\neq n$}, \\
 n!~{a_{1,1}^{2n+1-{a_{1,0}\over a_{2,0}}}}~~
a_{2,0}^{{a_{1,0}\over a_{2,0}}-2}
{(a_{1,0}+(n-1)a_{2,0}) \over(a_{1,0}+(2n-1)a_{2,0})}~\Gamma(1-n-{a_{1,0}\over a_{2,0}}),&\mbox{\quad if $m=n$.}
       \end{array} \right.
\end{align}
The proof of this identity follows by noting that
\begin{align*}
&\int\limits_{0}^\infty
{a_{1,1}^{n+m}\over a_{2,0}}~ {}_2F_0\left(-n,n-1+{a_{1,0}\over a_{2,0}};~-~
;-{a_{2,0}\over a_{1,1}}x\right)~ {}_2F_0\left(-m,m-1+{a_{1,0}\over a_{2,0}};~-~
;-{a_{2,0}\over a_{1,1}}x\right)e^{-{a_{1,1}\over a_{2,0}x}}x^{{a_{1,0}\over a_{2,0}}-2}dx\\
&={a_{1,1}^{n+m}\over a_{2,0}}\left({a_{1,1}\over a_{2,0}}\right)^{1-{a_{1,0}\over a_{2,0}}}\Gamma(1-{a_{1,0}\over a_{2,0}})
\sum_{j=0}^m
{(-m)_j(m-1+{a_{1,0}\over a_{2,0}})_j\over ({a_{1,0}\over a_{2,0}})_{j}j!}{}_2F_1(-n,n-1+{a_{1,0}\over a_{2,0}};{a_{1,0}\over a_{2,0}}+j;1)
\end{align*}
where we have used the series representation of the hypergeometric functions ${}_2F_0$ and ${}_2F_1$ as well as the integral representation of the Gamma function. Thus, using the identity ${}_2F_1(-n,\alpha;\beta;1)=(\beta-\alpha)_n/(\beta)_n$, we finally obtain
\begin{align*}
&\int\limits_{0}^\infty
{a_{1,1}^{n+m}\over a_{2,0}}~ {}_2F_0\left(-n,n-1+{a_{1,0}\over a_{2,0}};~-~
;-{a_{2,0}\over a_{1,1}}x\right)~ {}_2F_0\left(-m,m-1+{a_{1,0}\over a_{2,0}};~-~
;-{a_{2,0}\over a_{1,1}}x\right)e^{-{a_{1,1}\over a_{2,0}x}}x^{{a_{1,0}\over a_{2,0}}-2}dx\\
&={a_{1,1}^{n+m}\over a_{2,0}}\left({a_{1,1}\over a_{2,0}}\right)^{1-{a_{1,0}\over a_{2,0}}}{\Gamma(1-{a_{1,0}\over a_{2,0}})\over({a_{1,0}\over a_{2,0}})_n }
\sum_{j=0}^m
{(-m)_j(m-1+{a_{1,0}\over a_{2,0}})_j\over j!}
{(-1)^n(-j )_n\over ({a_{1,0}\over a_{2,0}}+n)_j}\\
&= n!{a_{1,1}^{2n+1-{a_{1,0}\over a_{2,0}}}}~~
a_{2,0}^{{a_{1,0}\over a_{2,0}}-2}
{(a_{1,0}+(n-1)a_{2,0})\Gamma(1-n-{a_{1,0}\over a_{2,0}}) \over(a_{1,0}+(2n-1)a_{2,0})}\delta_{nm}.
\end{align*}

In Table (\ref{tableI}), we present some examples of the classical differential equations and their classification.

\begin{table}[!h]%
\caption{Examples of Classical Differential Equations}
\label{tableI}\centering %
\begin{tabular}{ccccccc}
\toprule %
Name & &Differential Equation & &Case& &Interval\\ \toprule %
Hypergeometric&\quad\quad\quad &$x(1-x)y''+(c-(a+b+1)x)y'+n(n+a+b)y=0$&\quad\quad\quad & II&\quad\quad\quad &($0,1$)\\
Legendre/Spherical &\quad\quad\quad &$(1-x^2)y''-2xy'+n(n+1)y=0$&\quad\quad\quad  & III &\quad\quad\quad &(-1,1)\\
Chebyshev (first kind)&\quad\quad\quad &$(1-x^2)y''-xy'+n^2y=0$&\quad\quad\quad & III&\quad\quad\quad &(-1,1)\\
Chebyshev (second kind)&\quad\quad\quad &$(1-x^2)y''-3xy'+n(n+2)y=0$&\quad\quad\quad & III&\quad\quad\quad &(-1,1)\\
Hyperspherical/Gegenbauer&\quad\quad\quad &$(1-x^2)y''-2(1+k)xy'+n(n+2k+1)y=0$&\quad\quad\quad & III&\quad\quad\quad &(-1,1)\\
Jacobi&\quad\quad\quad &$(1-x^2)y''+(\alpha x+\beta)y'+[n(n-1)-n\alpha]y=0$&\quad\quad\quad & III&\quad\quad\quad &(-1,1)\\
Romanovski&\quad\quad\quad &$(x^2+1)y''+(\alpha x+\beta)y-[n(n-1)+\alpha n]y=0$&\quad\quad\quad & III&\quad\quad\quad &$(-\infty,\infty)$\\
Hermite&\quad\quad\quad &$y''-2xy'+2ny=0$&\quad\quad\quad & IV&\quad\quad\quad &$(-\infty,\infty$)\\
Laguerre&\quad\quad\quad &$xy''+(\alpha+1-x)y'+ny=0$&\quad\quad\quad & V&\quad\quad\quad &$(0,\infty)$\\
Confluent Hypergeometric&\quad\quad\quad &$xy''+(c-bx)y'+nby=0$&\quad\quad\quad & V&\quad\quad\quad &$(0,\infty)$\\
Bessel (Polynomials)&\quad\quad\quad &$x^2y''+((\alpha+2)x+\beta)y'+n(n+\alpha+1)y=0$&\quad\quad\quad & VI&\quad\quad\quad &$(0,\infty)$\\
\hline
\end{tabular}
\end{table}
\section{New solvable classes of differential equations}
\noindent{\bf Theorem 2.} \emph{For arbitrary differentiable function $Q(x)$, the differential equation
\begin{align}\label{eq69}
y_n''&+{Q(x)\over (a_{2,0}x^2+a_{2,1}x+a_{2,2})}y_n'\notag\\
&-\bigg({n(n-1)~a_{2,0}+n~a_{1,0}\over (a_{2,0}x^2+a_{2,1}x+a_{2,2})}+{1\over 2}{d\over dx}\left({-Q(x)+a_{1,0}x+a_{1,1}\over a_{2,0}x^2+a_{2,1}x+a_{2,2}}\bigg)+{(a_{1,1}+a_{1,0}x)^2-Q^2(x)\over 4(a_{2,0}x^2+a_{2,1}x+a_{2,2})^2}\right)y_n=0
\end{align}
has solutions given by 
\begin{equation}\label{eq70}
y_n(x)=f_n(x)\cdot \exp\left({{1\over 2}\int^x {-Q(t)+a_{1,0}t+a_{1,1}\over a_{2,0}t^2+a_{2,1}t+a_{2,2}}dt}\right),
\end{equation}
where the polynomials $f_n(x)$ are given explicitly by 
\begin{align}\label{eq51}
&f_n(x)={(-1)^n(\sqrt{a_{2,1}^2 - 4a_{2,0}a_{2,2}})^n
\left({2a_{2,0}a_{1,1}-a_{1,0}a_{2,1}-a_{1,0}\sqrt{a_{2,1}^2-4a_{2,0}a_{2,2}}\over -2a_{2,0}\sqrt{a_{2,1}^2-4a_{2,0}a_{2,2}}}\right)_n}\notag\\
&\times
{}_2F_1
\left(\begin{array}{ll}
-n & n-1+{a_{1,0}\over a_{2,0}} \\
{2a_{2,0}a_{1,1}-a_{1,0}a_{2,1}-a_{1,0}\sqrt{a_{2,1}^2-4a_{2,0}a_{2,2}}\over -2a_{2,0}\sqrt{a_{2,1}^2-4a_{2,0}a_{2,2}}}& ~ \\
\end{array}\bigg|{a_{2,0}x\over \sqrt{a_{2,1}^2-4a_{2,0}a_{2,2}}}+{a_{2,1}+\sqrt{a_{2,1}^2-4a_{2,0}a_{2,2}}\over 2\sqrt{a_{2,1}^2-4a_{2,0}a_{2,2}}}\right).
\end{align}
Note: if $Q(x)=a_{1,0}x+a_{1,1}$, the differential equation (\ref{eq69}) reduces to that of Theorem 1.}
\vskip0.1true in
\noindent{Proof:} Our proof of the theorem follows by analyzing the exact solutions of the differential equation
$$(a_{2,0}x^2+a_{2,1}x+a_{2,2})y''+Q(x)y'+(R(x)+\lambda)y=0$$
in the form
$y(x)=f(x)v(x)$, where $v(x)$ is the exact solution of the differential equation
$$(a_{2,0}x^2+a_{2,1}x+a_{2,2})v''+Q(x)v'+(R(x)+\mu)v=0,$$
which yields by direct substitution that
$$(a_{2,0}x^2+a_{2,1}x+a_{2,2})f''+{1\over v}[2(a_{2,0}x^2+a_{2,1}x+a_{2,2})v'+Q(x)v]f'+(\lambda-\mu)f=0,$$
where $\mu=\lambda+n(n-1)a_{2,0}+na_{1,0}$, if
$$2(a_{2,0}x^2+a_{2,1}x+a_{2,2}){v'\over v}+Q(x)=a_{1,0}x+a_{1,1}\Rightarrow v(x)=e^{{1\over 2}\int^x {-Q(t)+a_{1,0}t+a_{1,1}\over a_{2,0}t^2+a_{2,1}t+a_{2,2}}dt}$$
or
$${v'\over v}={-Q(x)+a_{1,0}x+a_{1,1}\over 2(a_{2,0}x^2+a_{2,1}x+a_{2,2})},{v''\over v}=\left({-Q(x)+a_{1,0}x+a_{1,1}\over 2(a_{2,0}x^2+a_{2,1}x+a_{2,2})}\right)^2+{d\over dx}\left({-Q(x)+a_{1,0}x+a_{1,1}\over 2(a_{2,0}x^2+a_{2,1}x+a_{2,2})}\right).
$$
The proof, then, follows by comparing the differential equation for $f(x)$ with that  given by Theorem 1.$\Box$
\vskip0.1true in

\noindent The following new classes of exactly solvable differential equations can be obtained by comparison with the special cases discuss in the previous section:
\begin{itemize}
  \item  The exact solution of the differential equation
\begin{align}\label{eq72}
&y_n''+{Q(x)\over (a_{2,1}x+a_{2,2})}y_n'-\bigg({n~a_{1,0}\over a_{2,1}x+a_{2,2}}+{d\over dx}
\left({-Q(x)+a_{1,0}x+a_{1,1}\over 2(a_{2,1}x+a_{2,2})}\bigg)+{(a_{1,1}+a_{1,0}x)^2-Q^2(x)\over 4(a_{2,1}x+a_{2,2})^2}\right)y_n=0
\end{align}
is given, for $n=0,1,2,\dots$, by 
\begin{equation}\label{eq73}
y_n(x)=f_n(x)\cdot \exp\left({{1\over 2}\int^x {-Q(t)+a_{1,0}t+a_{1,1}\over a_{2,1}t+a_{2,2}}dt}\right)
\end{equation}
where the polynomials $f_n(x)$ are given explicitly by 
\begin{align}\label{eq74}
f_n(x)&=a_{2,1}^n\left({a_{2,1}a_{1,1} - a_{2,2}a_{1,0}\over a_{2,1}^2}\right)_n{}_1F_1\left(-n;{a_{2,1}a_{1,1} - a_{2,2}a_{1,0}\over a_{2,1}^2}
;-{a_{1,0}\over a_{2,1}}x - {a_{2,2}a_{1,0}\over a_{2,1}^2}\right).
\end{align}
\item The exact solution of the differential equation

\begin{align}\label{eq75}
&y_n''+{Q(x)\over x(a_{2,0}x+a_{2,1})}y_n'-\bigg({n(n-1)~a_{2,0}+n~a_{1,0}\over x(a_{2,0}x+a_{2,1})}+{d\over dx}\left({-Q(x)+a_{1,0}x+a_{1,1}\over 2x(a_{2,0}x+a_{2,1})}\bigg)+{(a_{1,1}+a_{1,0}x)^2-Q^2(x)\over 4x^2(a_{2,0}x+a_{2,1})^2}\right)y_n=0
\end{align}
is given, for $n=0,1,2,\dots$, by 
\begin{equation}\label{eq76}
y_n(x)=f_n(x)\cdot \exp\left({{1\over 2}\int^x {-Q(t)+a_{1,0}t+a_{1,1}\over t(a_{2,0}t+a_{2,1})}dt}\right)
\end{equation}
where the polynomials $f_n(x)$ are given explicitly by 
\begin{align}\label{eq77}
f_n(x)&={(-1)^na_{2,1}^n
\left({a_{1,0}a_{2,1}-a_{2,0}a_{1,1}\over a_{2,0}a_{2,1}}\right)_n}
{}_2F_1(-n,n-1+{a_{1,0}\over a_{2,0}} ;{a_{1,0}a_{2,1}-a_{2,0}a_{1,1}\over a_{2,0}a_{2,1}};{a_{2,0}\over a_{2,1}}x+1).
\end{align}
\item For $a_{2,0}\cdot a_{2,2}<0$, the exact solution of the differential equation
\begin{align}\label{eq78}
y_n''&+{Q(x)\over (a_{2,0}x^2+a_{2,2})}y_n'\notag\\
&-\bigg({n(n-1)~a_{2,0}+n~a_{1,0}\over (a_{2,0}x^2+a_{2,2})}+{d\over dx}\left({-Q(x)+a_{1,0}x+a_{1,1}\over 2(a_{2,0}x^2+a_{2,2})}\bigg)+{(a_{1,1}+a_{1,0}x)^2-Q^2(x)\over 4(a_{2,0}x^2+a_{2,2})^2}\right)y_n=0
\end{align}
is given, for $n=0,1,2,\dots$, by 
\begin{equation}\label{eq79}
y_n(x)=f_n(x)\cdot \exp\left({{1\over 2}\int^x {-Q(t)+a_{1,0}t+a_{1,1}\over a_{2,0}t^2+a_{2,2}}dt}\right)
\end{equation}
where the polynomials $f_n(x)$ are given explicitly by 
\begin{align}\label{eq80}
f_n(x)&={(-2\sqrt{- a_{2,0}a_{2,2}})^n
\left({a_{2,0}a_{1,1}-a_{1,0}\sqrt{-a_{2,0}a_{2,2}}\over -2a_{2,0}\sqrt{-a_{2,0}a_{2,2}}}\right)_n}\notag\\
&\times
{}_2F_1
\left(\begin{array}{ll}
-n & n-1+{a_{1,0}\over a_{2,0}} \\
{a_{2,0}a_{1,1}-a_{1,0}\sqrt{-a_{2,0}a_{2,2}}\over -2a_{2,0}\sqrt{-a_{2,0}a_{2,2}}}& ~ \\
\end{array}\bigg|{a_{2,0}x+\sqrt{-a_{2,0}a_{2,2}}\over 2\sqrt{-a_{2,0}a_{2,2}}}\right).
\end{align}
\item The exact solution of the differential equation
\begin{align}\label{eq81}
y_n''&+{Q(x)\over a_{2,2}}y_n'
-\bigg({(2n+1)~a_{1,0}\over 2a_{2,2}}-{Q'(x)\over 2a_{2,2}}+{(a_{1,1}+a_{1,0}x)^2-Q^2(x)\over 4(a_{2,2})^2}\bigg)y_n=0
\end{align}
is given by 
\begin{equation}\label{eq82}
y_n(x)=f_n(x)\cdot \exp\left({{1\over 2 a_{2,2}}\int^x [-Q(t)+a_{1,0}t+a_{1,1}]dt}\right)
\end{equation}
where the polynomials $f_n(x)$ are given explicitly by 
\begin{align}\label{eq83}
f_n(x)&=(a_{1,0}x+a_{1,1})^n{}_2F_0\left(-{n\over 2},{-(n-1)\over 2};~-~
;{2a_{2,2}a_{1,0}\over (a_{1,0}x+a_{1,1})^2 }\right).
\end{align}
\item The exact solution of the differential equation
\begin{align}\label{eq84}
y_n''&+{Q(x)\over a_{2,1}x}y_n'-\bigg({n~a_{1,0}\over a_{2,1}x}+{d\over dx}\left({-Q(x)+a_{1,0}x+a_{1,1}\over 2a_{2,1}x}\bigg)+{(a_{1,1}+a_{1,0}x)^2-Q^2(x)\over 4(a_{2,1}x)^2}\right)y_n=0
\end{align}
is given by 
\begin{equation}\label{eq85}
y_n(x)=f_n(x)\cdot \exp\left({{1\over 2}\int^x {-Q(t)+a_{1,0}t+a_{1,1}\over a_{2,1}t} dt}\right)
\end{equation}
where the polynomials $f_n(x)$ are given explicitly by 
\begin{align}\label{eq86}
f_n(x)&=a_{2,1}^n\left({a_{1,1} \over a_{2,1}}\right)_n{}_1F_1\left(-n;{a_{1,1} \over a_{2,1}}
;-{a_{1,0}\over a_{2,1}}x \right).
\end{align}
\item For ${a_{1,0}\over a_{2,0}}<-(2n-1)$ and ${a_{1,1}\over a_{2,0}}>0$, the exact solution of the differential equation
\begin{align}\label{eq87}
y_n''&+{Q(x)\over a_{2,0}x^2}y_n'-\bigg({n(n-1)~a_{2,0}+n~a_{1,0}\over a_{2,0}x^2}+{d\over dx}\left({-Q(x)+a_{1,0}x+a_{1,1}\over 2a_{2,0}x^2}\bigg)+{(a_{1,1}+a_{1,0}x)^2-Q^2(x)\over 4(a_{2,0}x^2)^2}\right)y_n=0
\end{align}
is given by 
\begin{equation}\label{eq88}
y_n(x)=f_n(x)\cdot \exp\left({{1\over 2}\int^x {-Q(t)+a_{1,0}t+a_{1,1}\over a_{2,0}t^2}dt}\right)
\end{equation}
where the polynomials $f_n(x)$ are given explicitly by 
\begin{align}\label{eq89}
f_n(x)&=a_{1,1}^n~ {}_2F_0\left(-n,n-1+{a_{1,0}\over a_{2,0}};~-~
;-{a_{2,0}\over a_{1,1}}x\right).
\end{align}

\end{itemize}

\section{Conclusion}
Linear differential equations with non-constant coefficients represent an important class of problems.  If the  coefficients are powers equal  to the order of the derivative, then the analysis of Euler transforms the problem to a soluble equation with constant coefficients. If the coefficients are general polynomials with the same degree as the order of the derivative, no such general solution is known, even for the special case of second-order differential equations. What we offer in this paper is a complete analysis of those problems in this class that admit  exact polynomial solutions. We construct the solutions directly in terms of the parameters in the given polynomial coefficients. Where appropriate, we also discuss weight functions and orthogonality relations. This work provides a large collection of exact solutions which we hope will illuminate the spectrum of problems in mathematical physics that may be represented by the solutions of second-order linear differential equations.

\medskip

\section{Acknowledgments}

\medskip
\noindent Partial financial support of this work under Grant Nos. GP249507 and GP3438 from the 
Natural Sciences and Engineering Research Council of Canada
 is gratefully acknowledged by two of us (NS and RLH).

\section*{Appendix I: Note on the proof of Theorem 1}
\noindent In this appendix, we outline the proof of Theorem 1 using the asymptotic iteration method \cite{aim} that states the following. 
\vskip0.1true in
\noindent{\bf Theorem 2:} \emph{
Given $\lambda_{0}(x)$ and $s_{0}(x)$ in $C_{\infty}(a, b)$, the second-order linear homogeneous differential equation 
\begin{equation}\label{eq90}
y''=\lambda_{0}(x)y'+s_{0}(x)y
\end{equation}
has a general solution 
\begin{equation}\label{eq91}
y(x)=\exp(-\int^x\alpha(t)\,dt)\left[C_{2}+C_{1}\int^x \exp\left(\int^t(2\alpha(\tau)+\lambda_{0}(\tau))~d\tau\right)~dt\right].
\end{equation}
if, for sufficiently large $n>0$,  
\begin{equation}\label{eq92}
\frac{s_{n}}{\lambda_{n}}= \frac{s_{n-1}}{\lambda_{n-1}}\equiv \alpha(x)
\end{equation}
where
\begin{equation}\label{eq93}
\lambda_{k}= \lambda_{k-1}'+s_{k-1}+\lambda_{0}\lambda_{k-1}\quad \mbox{ and }\quad s_{k}=s_{k-1}'+s_{0}\lambda_{k-1}
\end{equation}
for $k=1, 2,...n$.
}
\vskip0.1true in
\noindent Initiating the asymptotic iteration method for the solution of equation (\ref{eq1})  with
\begin{equation}\label{eq94}
\lambda_0(x)= -{a_{1,0}x+a_{1,1}\over a_{2,0}x^2+a_{2,1}x+a_{2,2}}\quad\mbox{and}\quad 
s_0(x)={\tau_{0,0}\over a_{2,0}x^2+a_{2,1}x+a_{2,2}},
\end{equation}
and compute the sequences $\lambda_n$ and $s_n$ for $n\geq 1$. The termination condition $\delta_n\equiv \lambda_{n}s_{n-1}-\lambda_{n-1}s_n=0$, see equation (\ref{eq92}), then yields for $n=0,1,2,\dots$ (for $n=0$, we may compute $\delta_0$ starting with $s_{-1}=0$ and $\lambda_{-1}=1$), 
\begin{align*}
\delta_0&=0\Rightarrow \tau_{0,0}=0\notag\\
\delta_1&=0\Rightarrow \tau_{0,0}(\tau_{0,0}-a_{1,0})=0\notag\\
\delta_2&=0\Rightarrow \tau_{0,0}(\tau_{0,0}-a_{1,0})(\tau_{0,0}-2a_{2,0}-2a_{1,0})=0\notag\\
\delta_3&=0\Rightarrow \tau_{0,0}(\tau_{0,0}-a_{1,0})(\tau_{0,0}-2a_{2,0}-2a_{1,0})(\tau_{0,0}-6a_{2,0}-3a_{1,0})=0\notag\\
\delta_4&=0\Rightarrow \tau_{0,0}(\tau_{0,0}-a_{1,0})(\tau_{0,0}-2a_{2,0}-2a_{1,0})(\tau_{0,0}-6a_{2,0}-3a_{1,0})(\tau_{0,0}-12a_{2,0}-4a_{1,0})=0,
\quad\dots
\end{align*}
In general, we have
\begin{align}\label{eq95}
\delta_n&=0\quad\mbox{if and only if}\quad \prod_{k=0}^n \left(\tau_{0,0}-k(k-1)a_{2,0}-ka_{1,0}\right)=0.
\end{align}
Thus, the second-order differential equation (\ref{eq1}) has a polynomial solution of
degree $n$ if
$$\tau_{0,0}=n(n-1)a_{2,0}+na_{1,0},\quad n=0,1,2,\dots,$$
provided $a_{2,0}^2+a_{1,0}^2\neq 0$ which agree with equation (\ref{eq3}). The polynomial solutions can be evaluated explicitly, using (\ref{eq91}), namely
$$y_n(x)=\exp\left(-\int^x \alpha(t)dt\right)=\exp\left(-\int^x{s_{n-1}(t)\over \lambda_{n-1}(t)}dt\right),  
$$
which yields
\begin{align*}
y_0(x)&=1\notag\\
y_1(x)&=a_{1,0}x+a_{1,1}\notag\\
y_2(x)&= (a_{2,0}+a_{1,0})(a_{1,0}+2a_{2,0})x^2+2(a_{2,1}+a_{1,1})(a_{2,0}+a_{1,0})x+a_{1,1}(a_{2,1}+a_{1,1})+(a_{1,0}+2a_{2,0})a_{2,2}\notag\\
y_3(x)&=(a_{1,0}+2a_{2,0})(a_{1,0}+3a_{2,0})(a_{1,0}+4a_{2,0})x^3+3(a_{1,0}+2a_{2,0})(a_{1,0}+3a_{2,0})(2a_{2,1}+a_{1,1})x^2\notag\\
&+3(a_{1,0}+2a_{2,0})(3a_{1,1}a_{2,1}+2a_{2,1}^2+4a_{2,0}a_{2,2}+a_{1,0}a_{2,2}+a_{1,1}^2)x\notag\\
&+2a_{1,1}a_{2,1}^2+3a_{1,1}^2a_{2,1}+a_{1,1}^3+4a_{1,0}a_{2,1}a_{2,2}+10a_{2,0}a_{1,1}a_{2,2}+3a_{1,0}a_{1,1}a_{2,2}+12a_{2,0}a_{2,1}a_{2,2}
\end{align*}
and so on for $y_n$, $n\ge 4$. 	It is not difficult to show that these polynomials can be generated using equation (\ref{eq16}). Further, we note that these polynomial solutions satisfy the following recurrence relations for $n=2,3,\dots$:
\begin{align*}
y_2(x)&=\left({ (a_{1,0}+a_{2,0})(a_{1,0}+2a_{2,0})\over a_{1,0} }x+{(a_{2,0}+a_{1,0})(2a_{1,0}a_{2,1}+a_{1,0}a_{1,1}-2a_{2,0}a_{1,1})\over a_{1,0}^2}\right)y_1(x)\\
&+\left({(a_{1,0}+2a_{2,0})(a_{1,0}^2a_{2,2}-a_{1,1}a_{1,0}a_{2,1}+a_{1,1}^2a_{2,0})\over a_{1,0}^2}\right)y_0(x)\\
y_3(x)&=\left({ (a_{1,0}+3a_{2,0})(a_{1,0}+4a_{2,0})\over a_{1,0}+a_{2,0} }x+{(3a_{2,0}+a_{1,0})(4a_{1,0}a_{2,1}+a_{1,0}a_{1,1}+4a_{2,1}a_{2,0}-2a_{2,0}a_{1,1})\over (a_{1,0}+a_{2,0})(a_{1,0}+2a_{2,0})}\right)y_2(x)\notag\\
&+\left({2(a_{1,0}+4a_{2,0})(a_{1,0}^2a_{2,2}-a_{1,1}a_{1,0}a_{2,1}+a_{1,1}^2a_{2,0}+4a_{2,0}a_{1,0}a_{2,2}-a_{1,0}a_{2,1}^2+4a_{2,0}^2a_{2,2}-a_{2,0}a_{2,1}^2)\over (a_{1,0}+a_{2,0})(a_{1,0}+2a_{2,0})}\right)y_1(x).
\end{align*}
Thus, we may generalize this approach to a three-term recurrence relation as given by (\ref{eq15}). Equation (\ref{eq17}) follows directly from equation (\ref{eq16}) through the series representation of Gauss hypergeometric function ${}_2F_1$, equation (\ref{eq19}), and perform the limit operation as the factor $a_{2,1}^2-4a_{2,0}a_{2,2}$ approach zero. Equation (\ref{eq20}) follows from the Pearson differential equation (\cite{ismail}, page 635), namely
$${d\over dx}[(a_{2,0}x^2+a_{2,1}x+a_{2,2})w(x)]=(a_{1,0}x+a_{1,1}) w(x).$$


\end{document}